# Sub-eV uncertainties on gamma-ray energies using a magnetic microcalorimeter – Potential new reference energies

**Matias Rodrigues[1], Mostafa Lokman Zahir[1], Martin Loidl[1], Lucille Chambon[1], Quentin Drenne[1], Michael Müller[2], Sebastian Kempf[2,3], Etienne Nigron[4] and Ferid Haddad[4,5]**

[1]Université Paris-Saclay, CEA, List, Laboratoire National Henri Becquerel (LNE-LNHB), F-91120 Palaiseau, France
[2]Institute of Micro- and Nanoelectronic Systems (IMS), Karlsruhe Institute of Technology (KIT), Hertzstrasse 16, 76187 Karlsruhe, Germany.
[3]Institute for Data Processing and Electronics (IPE), Karlsruhe Institute of Technology (KIT), Hermann-von-Helmholtz-Platz 1, 76344 Eggenstein-Leopoldshafen, Germany.
[4]GIP ARRONAX, 1 rue Aronnax, CS10112, 44817 Saint-Herblain Cedex, France
[5]Laboratoire Subatech, UMR 6457, IMT Nantes Atlantique/CNRS-IN2P3/Nantes, France

E-mail: matias.rodrigues@cea.fr

## Abstract

We present new measurements of 27 gamma-ray energies ranging from 14 keV to 136 keV, obtained using magnetic microcalorimeters for energy-dispersive spectrometry. The spectrometer has eight pixels and achieves a high energy resolution between 15 eV and 30 eV across the entire energy range. It faces a cryogenic source sampler with four movable sources, each containing a mixture of radionuclides, including $^{169}$Yb and $^{57}$Co. Four gamma-ray lines from $^{169}$Yb and one from $^{57}$Co were used to calibrate the spectrometer and correct for its non-linearity in the energy range 14 keV – 130 keV. The gamma-ray energies, emitted through the decay of $^{57}$Co, $^{109}$Cd, $^{133}$Ba, $^{153}$Gd, $^{154}$Eu, $^{155}$Eu, $^{169}$Yb, $^{170}$Tm, $^{210}$Pb, $^{239}$Np, $^{241}$Am and $^{243}$Am, have been reassessed. This work significantly reduces the uncertainties upon previously reported gamma energies obtained by energy-dispersive spectrometry using semiconductor detectors: the lowest absolute uncertainty achieved is 0.13 eV at 105.3 keV, which corresponds to a relative uncertainty of 1.3 ppm. With respect to the available literature, among the 27 measured gamma-ray lines, the uncertainties of 10 energies were reduced by a factor of 5, while those of 4 additional energies were reduced by more than one order of magnitude.

## 1. Introduction

Precise gamma-ray energies below 200 keV with uncertainties well below 0.5 eV are needed to calibrate the energy scale of low temperature detectors (LTDs) and to correct their non-linearity [1–3]. Indeed, LTDs are energy-dispersive spectrometers that can achieve very high energy resolution below 100 mK. Resolving powers of about 5000 have been demonstrated, equivalent to a full width at the half-maximum ($FWHM$) energy resolution of 1.25 eV at 6 keV and 10 eV at 59.5 keV [1,4]. Such resolutions enable very precise measurements of gamma-ray energies because the contribution of the statistical uncertainty to the energy $E$ is given by the equation (1):

$$\sigma_{E,stat.}(E) = \frac{\sigma_{res.}(E)}{\sqrt{N_{count}(E)}} \quad (1)$$

where $\sigma_{res.}$ is the standard deviation of the Gaussian peak distribution, also given by the relation $FWHM = 2.35 \cdot \sigma_{res.}$, and usually limited by the detector resolution,
and $N_{count}$ is the number of counts in the peak of interest in the spectrum.

For this reason, LTDs are being developed and deployed in a range of applications to decrease the uncertainties on key quantities measured by high energy resolution photon spectrometry, such as line energies or emission intensities. Some of these applications relate to the measurement of X-ray energies emitted by exotic atoms with targeted uncertainties well below 0.5 eV [5–7]. Other applications are related to metrology in order to improve the databases, benchmark theoretical calculations of fundamental parameters, or improve materials analysis [8–12].

However, while the energies measured by LTDs are highly resolved, they are not accurate without precise energy calibration in order to both convert the histogram channel into an energy scale and correct the detector non-linearity. Indeed, LTDs are intrinsically non-linear due to their working principle described below. For photon spectrometry applications, most of the employed technologies are based on two types LTDs: magnetic microcalorimeters (MMCs) and transition edge sensors (TESs) [13,14]. These detectors consist of an absorber tightly thermally coupled with a thermal sensor. The energy $E$ deposited into the absorber by an incoming photon is converted to a temperature elevation $\Delta T$ that is, in a small signal approximation, given by:

$$\Delta T = E/C \quad (2)$$

where $C$ is the total heat capacity of the detector.
The detector remains linear as long as it is in the small signal regime, i.e. $\Delta T \ll T_0$, where $T_0$ is the detector temperature equilibrium.

The thermal sensor, acting as a thermometer, converts the temperature elevation $\Delta T$ into a change of another measurable physical quantity, i.e., a change of resistance $\Delta R$ for a TES or a change of magnetization $\Delta M$ for an MMC. The detector is weakly thermally coupled to the thermal bath by a weak thermal link to recover its equilibrium temperature $T_0$.

Achieving high energy resolution means to maximize the signal-to-noise ratio, the measurable signal being proportional to $\Delta T$ implies therefore minimising $C$. Then, high energy resolution means large $\Delta T$ that becomes non-negligible compared with $T_0$ and the pulse heights are no longer in the small signal regime; a non-linearity appears because the specific heats of the materials contributing to the heat capacity $C$ are temperature dependent. For LTDs, a high energy resolution is therefore inextricably linked to a non-linearity. Added to this, there is a second cause of non-linearity arising from the temperature dependence of the sensor sensitivities. For MMCs, the temperature sensitivity of the sensor magnetization, $\partial M/\partial T$, varies as $1/T$ [14]; whereas for TESs, the variation of resistance with the temperature, $\partial R/\partial T$, is specific to each TES, depending on the shape of the superconducting transition [15,16].

The non-linearity of LTDs can be corrected using the well-known gamma-ray and x-ray energies [8,10], which are recommended in the databases provided by the evaluations of Helmer and van der Leun for gamma-rays and Deslattes *et al.* for x-rays [17,18]. Regarding gamma-rays, the energies of the standards are evaluated from the experimental data obtained using two types of instrumentation: the wavelength-dispersive spectrometers (WDS) and the energy-dispersive spectrometers (EDS) with semiconductor detectors. WDS have a very high energy resolution and their measurements can be linked to the SI unit of the metre via the inter-reticular distance of the diffraction crystals obtained by optical interferometry [19]. However, they have a low detection efficiency and their measurements mainly focus on gamma-rays emitted by radionuclides with a relatively short half-life (less than a few hundred days) in order to avoid handling large quantities of radioactive materials. Two radionuclides stand out due to the very low uncertainties of the emitted energies: $^{161}$Tb and the $^{169}$Yb [20,21]. However, these radionuclides have short half-lives (6.89 (2) days and 32.018 (5) days respectively), making them unsuitable to calibrate LTDs during long time scale measurements. Furthermore, these radionuclides are difficult to purchase as they must be produced on demand. EDS is mostly used to measure the energies of radionuclides with longer half-lives. But, semiconductor detectors have a much poorer resolution and must be calibrated using other standard energies determined by WDS. Consequently, the uncertainties of the semiconductor-measured energies are greater than those obtained by WDS. Therefore, energy standards obtained by WDS are to be preferred.

This study aims to present new gamma-ray energy values below 200 keV with reduced uncertainties for commonly accessible radionuclides, and to provide metrological redundancy to validate previously reported energies derived from EDS using semiconductors. Some recommended energies published in the 2000 evaluation, obtained by Helmer himself, have indeed been questioned [22].

**Table 1.** Summary of the gamma-ray data of the radionuclides measured during this work. Energies with an asterisk * indicate the calibration lines $e_{cal}$,. Energies and decay data are from DDEP (Decay Data Evaluation Project) [23], excepted for the 14.4 keV gamma-ray energy emitted by $^{57}$Co, calculated as the mean value of measurements in the references [24–26] (see Section 4.2). The first column indicates which method (WDS or EDS) the recommended value originates from. The column [*Run; Source*] shows when (*Run*) and where (*Source $S_j$*) the radionuclides were measured in the present study. The *Radioactive solution* column describes the nature of the solution employed to produce the source.

| Method | Radionuclides | Decay mode | Half-life | Main energies[1] / keV | Intensities / 100 decays | Radioactive solution medium - carrier | [*Run; Source*] |
|---|---|---|---|---|---|---|---|
| WDS | $^{169}$Yb → $^{169}$Tm | EC | 32.018 (5) d | 63.12044 (4)*<br>93.61447 (8)*<br>109.77924 (4)*<br>118.18940 (14)<br>130.52293 (6)*<br>177.21307 (6) | 44.05 (24)<br>2.571 (17)<br>17.36 (9)<br>1.87 (1)<br>11.38 (5)<br>22.32 (10) | Irradiated foil of $^{169}$Tm | All |
| | $^{57}$Co → $^{57}$Fe | EC | 271.81 (4) d | 14.412492 (19)*<br>122.06065 (12)<br>136.47356 (29) | 9.18 (12)<br>85.49 (14)<br>10.71 (15) | HCl 0.1 mol/L - $CoCl_2$ | All |
| | $^{153}$Gd → $^{153}$Eu | EC | 240.4 (10) d | 69.67300 (13)<br>97.43100 (21)<br>103.18012 (17) | 2.42 (7)<br>29.0 (8)<br>21.1 (6) | HCl 0.5 mol/L - $GdCl_3$ | [R1&2; S1]<br>[R1; S4] |
| | $^{239}$Np → $^{239}$Pu | β⁻ | 2.356 (3) d | 61.460 (2)<br>106.125 (2) | 1.29 (2)<br>25.9 (3) | HCl - none | [R1&2; S4] |
| EDS | $^{109}$Cd → $^{109}$Ag | EC | 461.9 (4) d | 88.0336 (10) | 3.66 (5) | HCl 0.1 mol/L - $CdCl_3$ | [R1&2; S3] |
| | $^{133}$Ba → $^{133}$Cs | EC | 10.539 (6) a | 53.1622 (18)<br>79.6142 (19)<br>80.9979 (11) | 2.14 (6)<br>2.63 (19)<br>33.31 (30) | HCl 0.1 mol/L - $BaCl_2$ | [R1; S4] |
| | $^{155}$Eu → $^{155}$Gd | β⁻ | 4.753 (14) a | 26.531 (21)<br>45.299 (1)<br>60.0086 (10)<br>86.5479 (10)<br>105.3083 (10) | 0.316 (22)<br>1.31 (5)<br>1.22 (5)<br>30.7 (3)<br>21.1 (6) | HCl 0.5 mol/L - $EuCl_3$ | [R1&2; S2] |
| | $^{210}$Pb → $^{210}$Bi | β⁻ | 22.23 (12) a | 46.539 (1) | 4.252 (40) | $HNO_3$ 1 mol/L - none | [R1&2; S1] |
| | $^{241}$Am → $^{237}$Np | α | 432.6 (6) a | 26.3446 (2)<br>33.1963 (3)<br>59.5409 (1) | 2.31 (8)<br>0.1215 (28)<br>35.92 (17) | $HNO_3$ 1 mol/L - $EuCl_3$ | [R1&2; S1] |
| | $^{243}$Am → $^{239}$Np | α | 7.367 (23)×$10^3$ a | 43.53 (2)<br>74.66 (2) | 5.89 (10)<br>67.2 (12) | HCl 1 mol/L - none | [R1&2; S4] |

[1] The energies in keV unit, given the table 1, are based on the conversion factor wavelength to energy of 1.23984244×$10^{-6}$ eV·m, the same as used by Helmer and van der Leun in [17]

## 2. The experimental method and set-up

To benefit from both high energy resolution and reasonable detection efficiency, the energies of gamma-rays were measured by energy-dispersive spectrometry with an LTD. Among the LTD technologies, MMC-based LTDs were favoured due to their lower, smoother energy non-linearity, which follows a second- or third-order polynomial function of energy [14]. This characteristic reduces the number of calibration reference lines required to correct the non-linearity and diminishes both calibration uncertainties and potential systematic errors.

The radionuclides measured were selected on the basis of the recommended values of Helmer and von der Leun [17]. Their gamma-ray energies can be classified in two groups: those that serve to calibrate the MMC detector, and those to be reassessed that could potentially serve as future energy standards. The selected radionuclides, their characteristics and their main emitted energies are summarised in table 1. The calibration radionuclides were chosen based on the following criteria: energies below 200 keV, measured by WDS with uncertainties below 0.2 eV, having reasonably large emission intensities (> 1 per 100 decays), and having half-lives longer than 10 days. The radionuclides selected to reassess their gamma-ray energies must be commercially available and have half-lives greater than a few hundred days. They must also emit gamma-rays below 200 keV with also reasonably large emission intensities larger than 1 per 100 decays. The radionuclides selected to correct the detector non-linearity or to check the energy calibration are $^{57}$Co, $^{153}$Gd and $^{169}$Yb (table 1). Cobalt-57 is of particular interest for calibration purposes as it emits a gamma ray at a very low energy of 14.4 keV. Ytterbium-169 emits many gamma rays between 60 keV and 200 keV having energies with very low uncertainties. Moreover, $^{169}$Yb was preferred to $^{161}$Tb, because in addition to having a longer half-life, it provides an absolute measurement linked to the SI unit of wavelength through the interstitial lattice spacing of the spectrometer crystal [19,20]. In contrast, the gamma rays of $^{161}$Tb were measured relative to the $^{170}$Tm gamma-ray energy at 84 keV also measured in reference [20]. This relative measurement also leads to larger uncertainties on the recommended values of $^{161}$Tb.

A dedicated experimental setup has been designed to measure selected radionuclides. It comprises two mechanical stages: one supports the MMC spectrometer and the other supports a cryogenic source sampler. The MMC spectrometer was designed using MMC chips that were microfabricated on a silicon wafer (figure 1) [27]. Each chip contains several layers corresponding to different materials and functionalities. The paramagnetic sensor is made of silver doped with 475 ppm of enriched erbium-168. The paramagnetic sensor's change in magnetisation is read out by a planar, meander-shaped pick-up coil. Two of these pick-up coils, made of niobium, are connected in parallel to form, below 9.3 K, a closed superconducting circuit through which a permanent current circulates. This creates a local magnetic field that magnetises the sensor covering the pick-up coils. These are connected to the input coil of a two-stage superconducting quantum interference device (SQUID) sensor, which converts changes in magnetic flux in the input coil into changes in voltage with a very low noise level. In addition to the niobium electrical bond pads used to connect the coils and inject the permanent current, a large gold bond pad on the chip thermally connects the chip to the thermal bath via gold wires (Figure 1). The sensor is connected to this thermal gold pad via a thermal link formed by a gold film strip dimensioned to produce pulses with a decay time constant of around 1 ms at 20 mK.

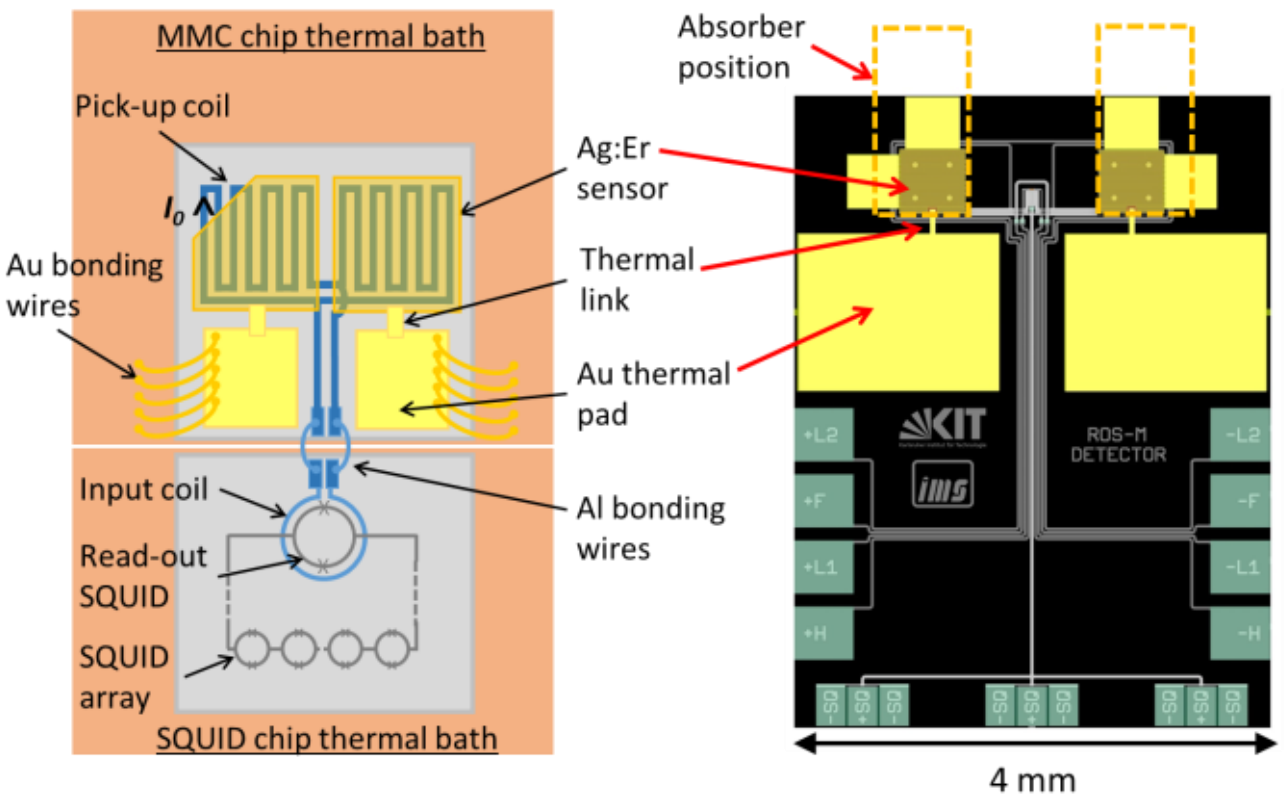

**Figure 1.** On the left, the scheme of the MMC chip connected to the two-stage SQUID chip, both sitting on their respective thermal baths. On the right, the scheme of the final MMC chip layout. For details, see also [27]. The dashed lines represent the position of the rectangular gold absorbers.

To increase the total detection efficiency without compromising energy resolution, the MMC comprises eight gold-based absorber pixels. Each absorber pixel is a rectangular parallelepiped with an area of 0.75×1.5 mm² and a thickness of 50 µm. The absorbers were glued to the sensors of four MMC chips using Stycast 1266 epoxy [27]. Each MMC chip is read out by a two-stage SQUID sensor. Two absorbers on the same MMC chip produce pulses of opposite polarity in the input coil of the SQUID.

The MMC and SQUID chips sit on two separate copper parts, which act as thermal baths to prevent power dissipation from the SQUID from increasing the temperature of the MMC sensors. The copper holders and chips are enclosed in a lead superconducting electromagnetic shield with an opening above the MMCs to allow photons to pass through towards the absorbers. A 1-mm-thick beryllium window is positioned in front of the absorbers to block alpha particles and absorb most of the electrons emitted during radionuclide decay.

The stage holding the cryogenic source sampler comprises a rotary motor with a copper arm supporting a tray containing four source slots (figure 2). Using a source sampler has several advantages: the sources can be changed while maintaining the same working conditions for the detectors, SQUIDs and cryostat. Consequently, the detector exhibits practically identical non-linearity with each source. At the same time, it enables the sources to be positioned as close as possible to the absorbers inside the cryostat, thereby maximising detection efficiency and minimising the required activity for measurement. It also reduces spectral interference between energy peaks by placing the appropriate radionuclides in separate source slots. The motor is driven by piezoelectric elements and can rotate endlessly. Its position is read by an optical encoder and fibre with an accuracy of approximately 0.43°. Switching between two sources takes around six hours to prevent the cryostat from heating above 400 mK.

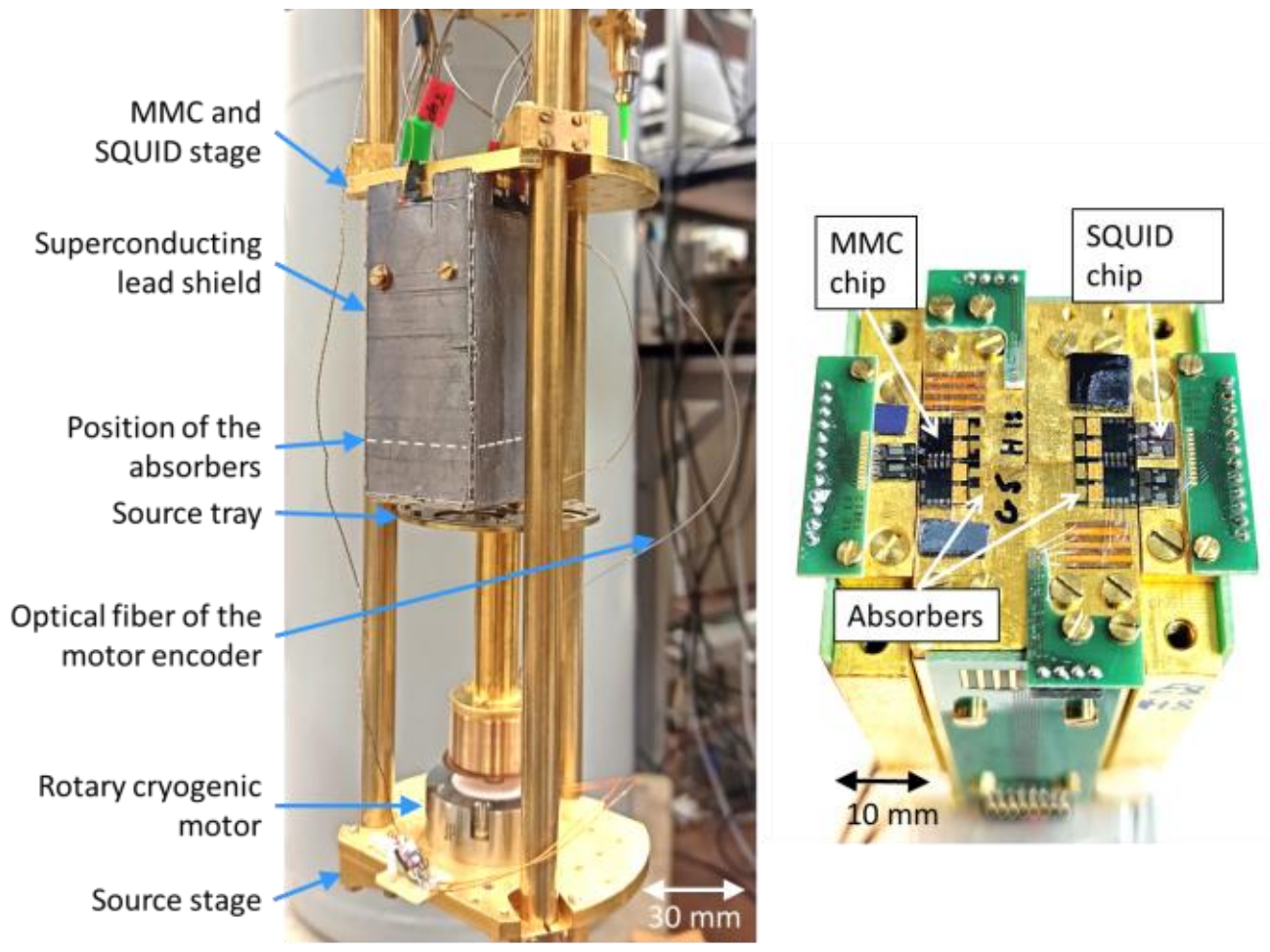

**Figure 2.** On the left, photograph of the two experimental stages mounted on the mixing chamber of the wet dilution refrigerator. On the right, photograph of the MMC spectrometer mounting with its different chips.

Each of the four sources contains a mixture of several radionuclides, which are listed in table 1. At least two of these radionuclides are dedicated to energy calibration: $^{57}$Co, $^{169}$Yb and/or $^{153}$Gd. These mixtures were chosen to minimise spectral interference between the peaks of interest; their spectra as they would occur in the detector were calculated using Monte Carlo simulations, in particular for peak interference due to detector artefacts like photon escape.

The sources were prepared from radioactive solutions, the compositions of which are given in table 1. All the radionuclides except $^{169}$Yb were available in the laboratory or commercially. Ytterbium-169 was produced at GIP ARRONAX using a C70XP cyclotron [28]. A foil of $^{169}$Tm with a thickness of approximately 100 µm was exposed to a uniform deuteron beam with an energy of 14.4 MeV and an intensity of 40 nA for a duration of 1 hour and 15 minutes [22]. The impurity activities were determined by conventional gamma-ray spectrometry, the most dominant radionuclides, with half-lives longer than $^{169}$Yb, are $^{168}$Tm decaying to $^{168}$Er, $^{182}$Ta decaying to $^{182}$W and $^{170}$Tm decaying to $^{170}$Er. In the MMC spectrum, only two gamma-ray lines (79.8 keV from $^{168}$Tm and 84.3 keV from $^{170}$Tm) and two K$\alpha$ x-ray lines (48.2 keV and 49.1 keV from Erbium) were detected without interfering with other lines. As $^{169}$Yb must be present in all energy spectra, a piece of irradiated thulium with the desired activity was cut out and fixed to the beryllium window.

Two measurement sessions were carried out under different conditions, the conditions for the second session being updated based on feedback from the initial session. These sessions are subsequently referred to as R1 and R2. During the two sessions, the output voltage of the SQUID electronics was recorded continuously after amplification and application of a low-pass filter with a cut-off frequency of 100 kHz to prevent aliasing. The data were digitised at a sampling frequency of 500 kHz with a resolution of 16 bits using two Innovative Integration P25M cards, each with two input channels.

During the first session (R1), we encountered several issues:

- Only four of the eight pixels were working.
- The DC voltage levels of the SQUID output shifted abruptly into a range where the amplifier was no longer linear for certain pixels. This occurred despite the temperature of the unregulated dilution refrigerator being relatively stable at around 16.5 mK.
- For one pixel, the slew rate of the SQUID electronics was too slow to follow pulses larger than 100 keV, resulting in a change of the SQUID working point. The corresponding energies were unusable for calibration, resulting in poor non-linearity correction. This was due to imperfect electromagnetic shielding of the SQUID array in the preamplifier SQUID stage, which reduced the V-Φ (voltage vs. magnetic flux) characteristic.

These problems were solved by applying some changes during session R2. The temperature of the detector stage was regulated to 17 mK in order to maintain the voltage level within a narrower range and keep the amplifiers within the linear range. Although the baseline voltage variations were smaller, they were faster, which leads to a degradation in resolution for peaks at energies above 50 keV. The magnetic shielding of the MMCs and SQUIDs was improved by reducing the aperture of the lead shield and adding a high-permeability shield. Six of the eight pixels were operational.

For each session, data were recorded over a period of two weeks, i.e. approximately 3.5 days per source. Following analysis, the spectra had $4.6\times10^6$ and $12.3\times10^6$ counts respectively for the first and second sessions.

## 3. Detector characterization and performances

### *3.1 Pulses, noises and energy resolutions*

The detector and its four working pixels were characterised during the session R1; the second session showed very similar performance in terms of MMC sensitivity, noise, and energy resolution. Figure 3 shows the averaged and normalised pulses from each of the four operating absorbers, expressed in units of $\Phi_0$/keV. These pulses $p(t)$ were perfectly fitted by a two-exponential function given by equation (3).

$$p(t) = A\left(e^{-(t-t_0)/\tau_d} - e^{-(t-t_0)/\tau_r}\right) \qquad (3)$$

where $A$ is the sensitivity in units of $\Phi_0$/ keV,
$t$ is the time,
$t_0$ is the time at which the energy is deposited in the absorber,
$\tau_r$ and $\tau_d$ are respectively the rise and the decay time constants expressed as a fraction of 1/$e$ of the pulse height.

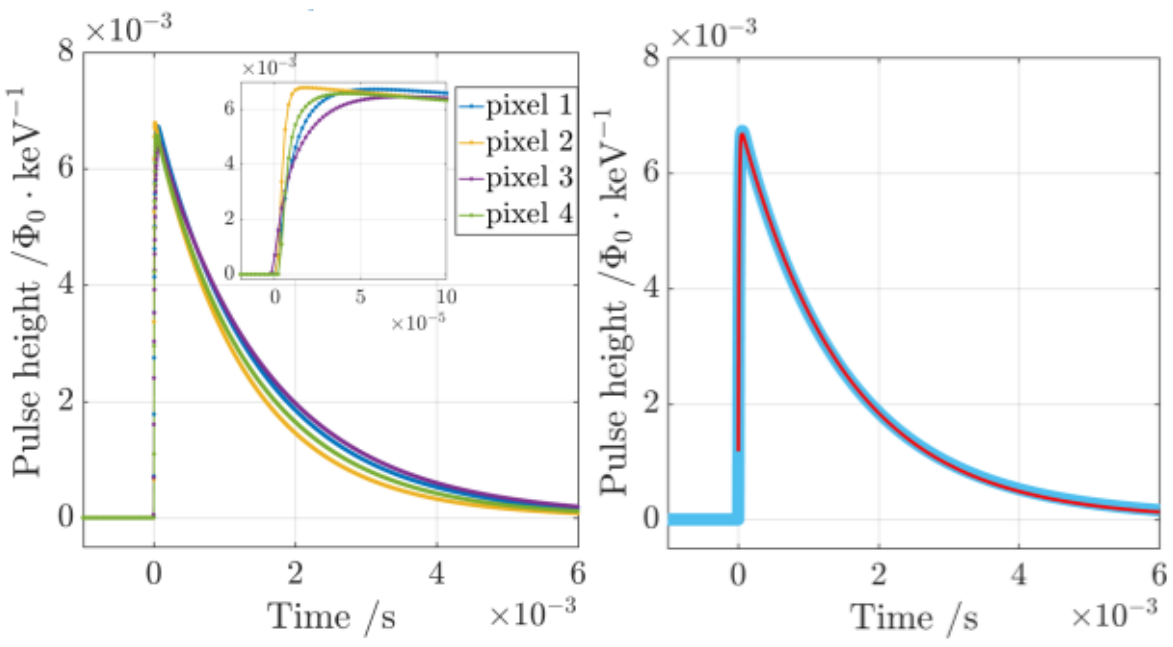

**Figure 3.** Left, the averaged pulses of the four operating absorbers of the first session. The inset shows a closer view of the pulse rise. Right, the averaged pulse of one of the absorbers (blue curve) fitted with a two-exponential function (red curve).

These fitting parameters are summarised in Table 2. Figure 3 shows that the pulses have very similar characteristics, except for the rise time, $\tau_r$. The latter varies because the absorbers are glued to the sensor, which is a non-reproducible method; nevertheless, this parameter has little impact on the energy resolution. The slight variability in sensitivity ($A$) is mainly caused by differences in absorber mass of around 10%. The variability in the constant $\tau_d$ is likely to be dominated by uniformity in the microfabrication process over the wafer surface. The calculated sensitivity, using the parameters of the microfabricated superconducting flux transformer and the thermodynamic properties of the MMC materials, as well as some operating parameters, is 6.81 m$\Phi_0$/keV, in very good agreement with the experimental ones *[29,30]*.

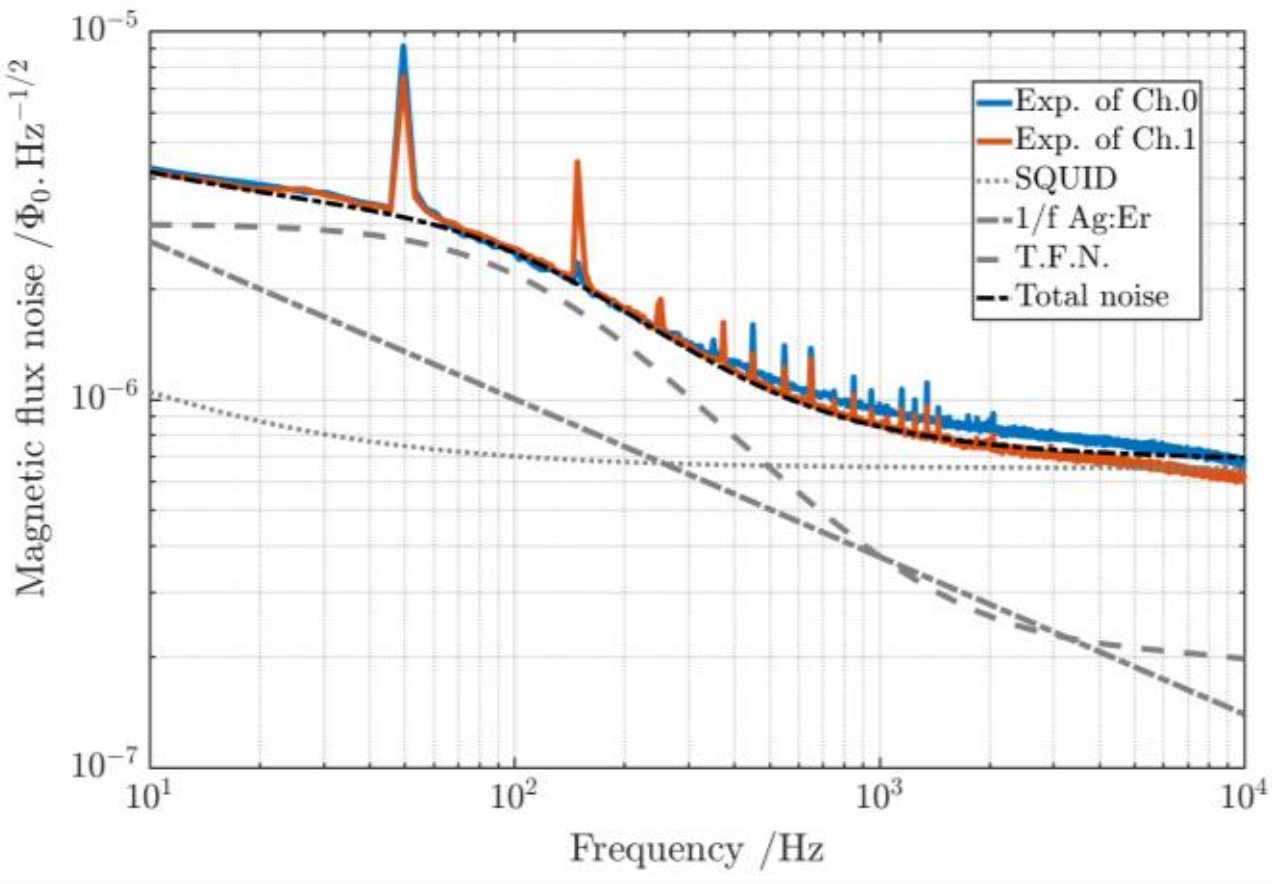

**Figure 4.** Comparison between the experimental noise spectral density of the magnetic flux (solid lines), which was measured at the readout of two SQUID channels, with the calculated noise (black dashed line). The calculated noise consists of several contributions (grey lines). The grey dotted line represents the SQUID noise; the dashed-dotted lines correspond to the 1/f noise of the Ag:Er sensor; and the grey dashed line shows the contribution from thermodynamic fluctuations in the internal energies of the MMC.

In addition to the signal, the measured noise was compared with the expected level of noise. The total noise comprises the noise from the SQUID and its electronics, as well as the noise from the detector (figure 4). Of these, the white noise level of the SQUID electronics was higher than expected, by about a factor of two, at 0.65 µ$\Phi_0$/Hz$^{1/2}$. This led to a 20% degradation in the expected resolution. Taking this extra noise level into account, we observed good agreement between the experimental and calculated noise spectral densities.

**Table 2.** Characteristics of the four operating absorbers during the session R1.

| Pixel of R1 | Sensitivity $A$ / m$\Phi_0$/keV | $\tau_d$ / ms | $\tau_r$ / µs | $\Delta E_{cal}$ / eV | $\Delta E_{exp}$ / eV |
|---|---|---|---|---|---|
| 1 | 6.98 | 1.50 | 10.7 | 15.0 | 14.8 |
| 2 | 6.82 | 1.29 | 2.33 | 15.1 | 15.8 |
| 3 | 6.79 | 1.62 | 15.2 | 15.2 | 15.3 |
| 4 | 6.72 | 1.42 | 5.41 | 14.9 | 16.2 |

The calculated and the expected experimental full width at half maximum (FWHM) energy resolutions, respectively $\Delta E_{cal}$[2] and $\Delta E_{exp}$, were compared. The calculated resolution $\Delta E_{cal}$ is based on the modelled signal responsivity and the modelled total noise spectral density derived from the detector parameters and the experimental working conditions *[30]*. On the other hand, the experimental resolution $\Delta E_{exp}$ is calculated based on the measured signal responsivity and the measured total noise spectral density. Table 2 summarises these energy resolutions for each absorber of session R1. These resolutions are very close to each other, meaning that the MMC performances can be predicted with confidence, it also means that the 50 Hz signal and its harmonics observed in figure 4 have little impact on the resolution.

### *3.2 Pulse height analysis*

[2] For the calculation of $\Delta E_{cal}$ some experimental parameters have been included (time constants, absorber mass and SQUID white noise) [30,31].

The recorded streams of data for each channel contain both positive and negative pulses that correspond to the two absorbers of one MMC chip. The arrival times of the pulses are identified using a band-pass filter and both positive and negative triggers. An extendable dead time is then applied, and the height of the pulses arriving during the live time is estimated using two filtering methods. The first method uses an optimal filter (OF), while the second uses a template fit (TF) scaled on the pulse using the least squares method. The energy resolutions obtained at peaks below 30 keV (Figure 5) are very close to the expected resolution $\Delta E_{exp}$. However, while the energy resolution degrades slowly with energy for R1, it increases quickly for R2 using the OF. This degradation increases the statistical uncertainty of the measured energies. To maintain a targeted uncertainty of $\sigma_E \leq 0.1$ eV, the counting statistics would need to be increased significantly, as shown in Figure 5 and on the right-hand y-axis. The reason for the resolution degradation during R2 is the temperature regulation using a PID controller. While this allows the DC voltage level to be kept within a range of less than 0.2 V, well below the full range of the preamplifier (3 V), it introduces relatively fast temperature variations within this range, with a period of about 30 s which causes the pulse baselines to become unstable. In the frequency domain, this manifests as low-frequency components that act as non-stationary noise for the OF, meaning that the OF is therefore no longer optimal for estimating the pulse amplitudes.As an alternative, the pulse heights of R2 were estimated using the TF method. Figure 5 shows that the resolution degradation is mitigated because the slope of the pre-trigger baseline was corrected by applying a linear fit to the pre-trigger pulse baseline, and because fewer pulse samples were used (768 vs. 8,192 for the OF). Using the TF method for R2 gives a resolution with a similar energy dependency to that of R1 with OF, although the resolution is slightly worse.

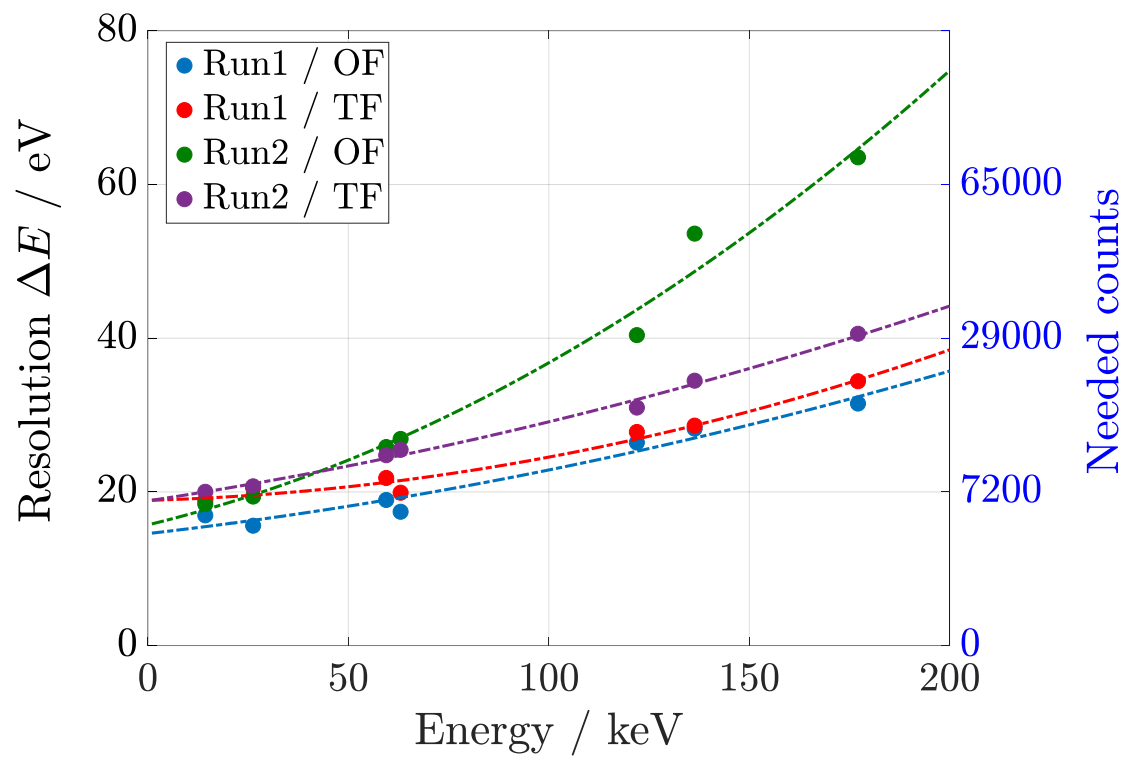

**Figure 5.** Energy resolution as the full width at half maximum as a function of the energy with two methods (OF and TF) used to estimate the pulse height. The right y-axis shows the required counting statistics to achieve a statistical uncertainty contribution of 0.1 eV.

### *3.3 Pulse height drift correction*

There is a drift of the pulse height as a function of time due to the temperature fluctuations of the thermal bath. It is mandatory to correct them to reach the resolutions shown in figure 5. For R1, the drift was corrected by fitting a spline function to the estimated pulse height of the 122 keV events ($^{57}$Co) as a function of time. Then the spline function was applied to all the pulse heights as a weight to correct the drift.

This method cannot be applied for R2 because the period of the temperature drift is comparable to the inverse of the 122 keV pulse rate per pixel (~ 30 s), which is insufficient for sampling the temperature drift. In this case, the linear correlation between the pre-trigger baseline level and the pulse height was used to establish a drift correction; indeed, the pre-trigger baseline level as well as the pulse height both vary linearly with the bath temperature at the first order. For this session, the drift correction was established from the baseline level for the 63 keV events.

### *3.4 Pulse selection and rejection*

The final step in processing the pulse data involves rejecting pile-up and anomalous pulses, which can introduce artefact spectral features into the energy spectra. This is achieved through pulse shape discrimination, whereby a full template pulse is scaled to the triggered pulses using a least squares method, similar to the TF method described above. The resulting mean square error (MSE) is then calculated. Figure 6 shows the estimated pulse energy by OF as a function of $MSE_{\mathrm{FTF}}$. Several pulse populations can be clearly distinguished; lower $MSE_{\mathrm{FTF}}$ values indicate that the pulse shape conforms to the model. Population 1 comprises the regular pulses selected for measuring photon energies and represents about 60% of the triggered events. The other populations displayed in Figure 6 are rejected.

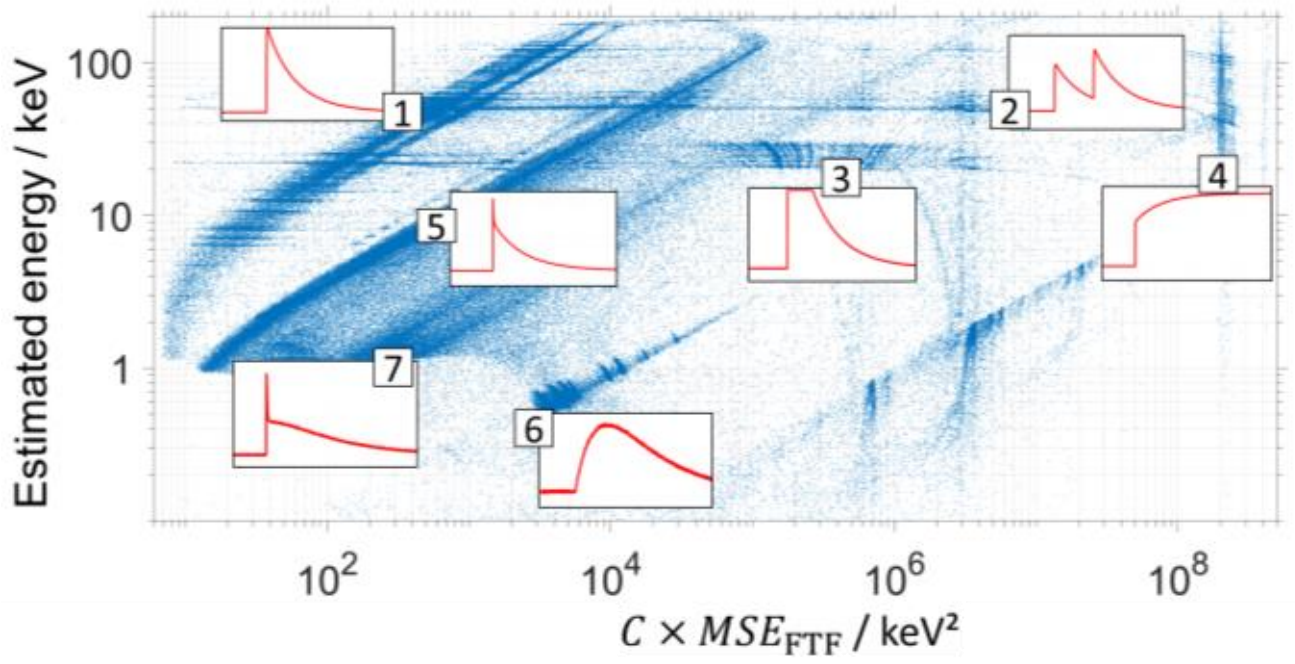

**Figure 6.** Estimated energy of the detected pulses as a function of the $MSE_{\mathrm{FTF}}$, $C$ is a scaling constant. Some pulse populations are identified as follow. Populations 1 to 4 involve photon hits in the absorber: 1) regular pulses to be selected, 2) pile-ups between two regular pulses, 3) ADC saturations, 4) change of the SQUID working point. Populations 5 to 7 involve photon hits outside the absorber: 5) sensor hits, 6) hits in the silicon substrate of the MMC chip, 7) supposed to be hits in the silicon substrate underneath the sensor.

## 4 Energy calibration and non-linearity correction

### *4.1 energy estimation*

Once the pulse heights have been determined and the regular events selected, the temperature drifts are corrected and a linear calibration is performed by multiplying the pulse heights by the tabulated energy of the events used for drift correction: 122.06065(12) keV and 63.12044(4) keV for R1 and R2, respectively. The pulse energy after linear calibration is called $e_{lin}$.

The entire set $\epsilon$ of pulse energies consists of 40 datasets $\epsilon_{i,j}$ such as $\epsilon = \{\epsilon_{1,1}, \dots, \epsilon_{P,S}\}$, where $P$ is the total number of working pixels ($i$ = 1, …, 4 pixels for R1 and $i$ = 5, …,10 for R2) and $S$ = 4 is the number of sources. Each dataset $\epsilon_{i,j}$ contains a number $M_{i,j}$ of pulse energies $e_{lin,i,j,k}$:

$$\epsilon_{i,j} = \{e_{lin,i,j,1}, e_{lin,i,j,2}, \dots, e_{lin,i,j,M_{i,j}}\}.$$

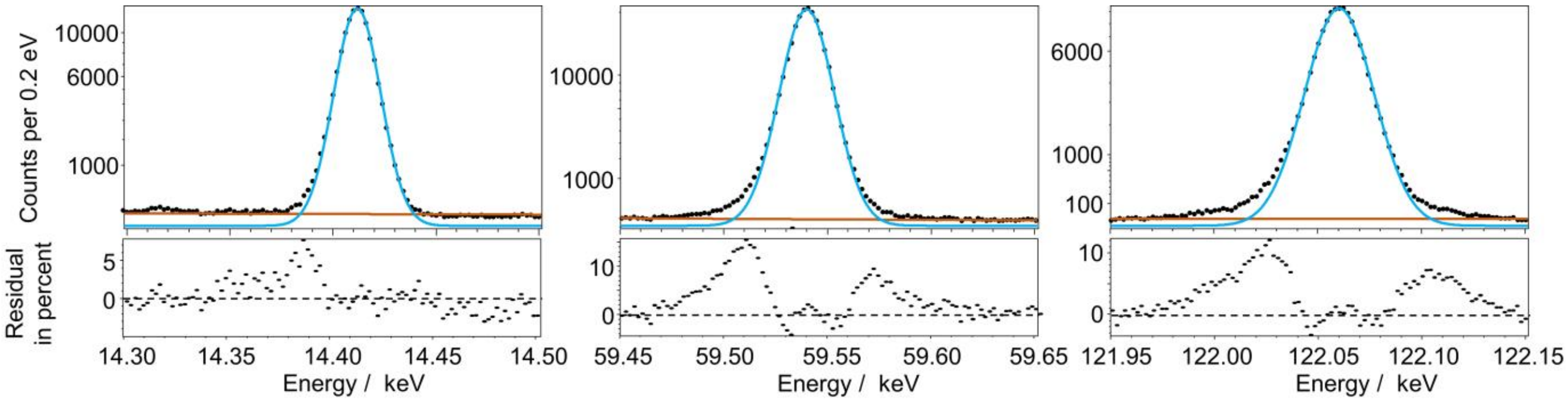

**Figure 7.** Histograms of gamma rays at three energies, with a square-root scaling of the Y-axis. The blue line shows the fit of a Gaussian function and the red line shows the fit of the background with a linear function. The histograms at the bottom show the corresponding residuals between the experimental data and the fit.

For each dataset $\epsilon_{i,j}$, the peak energy must be estimated using the pulse energies $e_{lin,i,j,k}$ forming the peak. In energy-dispersive spectrometry, the standard approach is to analyse the energy spectrum by fitting a region of interest (ROI) in the histogram with representative functions. This method has the advantage of distinguishing the peaks of interest from other spectrum features, or distinguishing overlapping peaks, by choosing the appropriate functions. It is ideal when the spectrometer exhibits a nearly Gaussian response function at full energy peaks (FEPs), in which case the peaks can be fitted with a Gaussian function for gamma lines.

In the MMC spectra, however, the peaks have asymmetric tails on both sides of the centroid. These tails may have multiple origins, such as an overcorrection of temperature fluctuations, thermal crosstalk between pixels on the same chip or thermal crosstalk between the substrate and the sensors [32]. Since the source is not collimated, many photons may interact with the substrate or with microfabricated structures of the MMC chips near the sensors.

Figure 7 shows three gamma-ray peaks in different energy regions that have been fitted with Gaussian functions. The resulting residuals clearly show asymmetric tails. Furthermore, the spread and intensity of the tails increase with energy. Using a Gaussian function with two tails greatly reduces the residuals. However, the correlation between the fitting parameters and their variances increases. Each tail is indeed an exponential function with two fitting parameters representing its amplitude and slope. Therefore, seven parameters are required to fit a single gamma-ray peak, as opposed to three when using a simple Gaussian. We observed that the centroid of the peaks shifts by a few hundred meV depending on whether the peak is fitted using a Gaussian or a two-tailed Gaussian function. These shifts are not accounted for in the estimated uncertainty, which could result in systematic errors.

As an alternative, a more basic processing method was used to determine the peak energies. Given the MMC's high energy resolution, peak interferences are rare. Thus, processing the spectrum by fitting the function is not essential for most gamma-ray lines. The basic method involves selecting a narrow region of interest (ROI) around a peak without interferences. The mean value $\bar{e}_{ROI,i,j}$ and the standard deviation $\sigma_{\bar{e}_{ROI,i,j}}$ of the pulse energies present in the ROI are then calculated. Pulse energies, $e_{lin,i,j,k}$, that do not respect the criteria of equation (4) are rejected.

$$\left|e_{lin,i,j,k} - \bar{e}_{ROI,i,j}\right| < C_{cut} \cdot \sigma_{\bar{e}_{ROI,i,j}} \qquad (4)$$

where $C_{cut}$ is a cut-off parameter used to reject energies that are far from the mean value, $\bar{e}_{ROI,i,j}$ , i.e. the peak centroid.

Then, the median $\tilde{e}_{lin,i,j} = med\left(e_{lin,i,j,k}\right)$ is calculated for the remaining number of events $e_{lin,i,j,k}$. Using the median to estimate the energy mitigates the contribution of the tails and background by excluding outliers. Figure 8 shows the obtained energies $\tilde{e}_{lin,i,j}$ for three different gamma-ray lines as a function of $C_{cut}$. It also compares the energies when the mean value (blue) is used instead of the median (red). The median energies, $\tilde{e}_{lin,i,j}$, are less sensitive and less dependent on $C_{cut}$ than the mean values.

However, using the median introduces increased statistical uncertainty, with a standard deviation approximately 1.253 times greater than that of the mean for a Gaussian distribution[3]. However, a larger $\sigma_{cut}$ value can be used for the median compared to the one used for the mean, taking a larger number of events $N_{cut}$ into account. A value of $C_{cut} = 2$ was used. For the gamma-ray lines, which have a distribution close to a Gaussian, the statistical uncertainty contribution, $\sigma_{stat,\tilde{e}_{lin,i,j}}$, associated with the median energies $\tilde{e}_{lin,i,j}$, is given by:

$$\sigma_{stat,\tilde{e}_{lin,i,j}} \approx \frac{1.253 \times \sigma_{\bar{e}_{cut,i,j}}}{\sqrt{N_{cut}}} \qquad (5)$$

where $\sigma_{\bar{e}_{cut,i,j}}$ is the standard deviation of the selected energies $e_{lin,i,j,k}$.

A Type B uncertainty, $\sigma_{ROI}$, was estimated from the dispersion of the data in Figure 8. This is related to the dependence of the energy on the choice of $C_{cut}$. $\sigma_{ROI}$ was estimated to be $3.5\times10^{-2}$ eV and was chosen as energy independent.

[3] The factor $\frac{1}{\sqrt{2/\pi}} = 1.253$ is applicable when the median is calculated for large samples of normally distributed data.

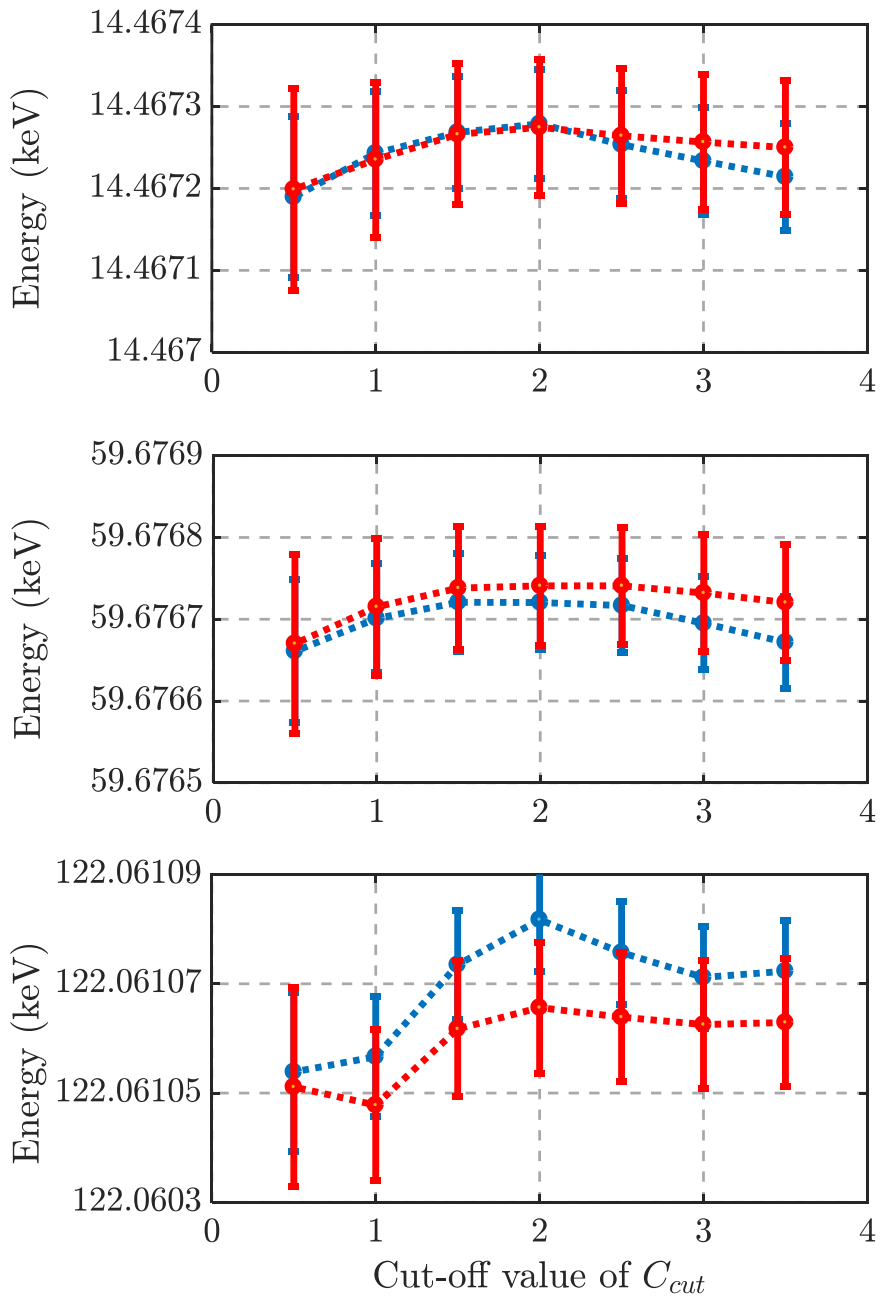

**Figure 8.** Estimated energy of the pulse events, $e_{lin,i,j,k}$, present in the ROI of different gamma-ray lines and satisfying the condition of equation (4) as a function of the cut-off parameter $C_{cut}$. In red data, the energy $\tilde{e}_{lin,i,j}$ is estimated using the median of the pulse energies $e_{lin,i,j,k}$. In blue data, the energy is estimated using the mean of the pulse energies $e_{lin,i,j,k}$.

## *4.2 Non-linearity correction*

To correct the non-linearity, we opted for four gamma lines of $^{169}$Yb and for the 14 keV gamma-line of $^{57}$Co, all of them were obtained by WDS experiments linked to the SI unit of meter. The 14 keV gamma-ray line appeared particularly useful, as it provides a calibration point in the low-energy range where only few gamma-ray lines are available. The recommended value of the 14 keV gamma line is derived from the difference between the 136 keV and 122 keV measurements made by Borchert, i.e. 14.41295 (31) keV [33]. However, this value does not align with three more recent and precise measurements [24–26], in which the wavelength was measured with an accuracy of up to 0.19 pm (i.e. 2.7 meV) by directly comparing it with the lattice spacing of a silicon single crystal reference. Despite their results being inconsistent because of very low uncertainties, they have a maximum deviation of only 19 meV between each other, and they are lower by about 420 meV than the recommended value. Therefore, we decided to take the calibration energy at 14 keV as the mean value of the three measurements with an uncertainty corresponding to the maximum deviation within the dataset, i.e. 14.412492 (19) keV . This uncertainty has no real impact on the calibration because, the corresponding weight of this line will be dominated by the uncertainties $\sigma_{ROI}$ and $\sigma_{stat,\tilde{e}_{lin,i,j}}$ (see equation (6)).

Despite the fact that K$\alpha$ x-ray energies are well known and intense in the obtained spectra, the x-rays were not used for the non-linearity correction for the following reasons:

- x-rays have a Voigt profile with an energy-dependent natural line width that dominates the detector Gaussian width above 20 keV,
- x-ray energies depend on the chemical environment; most of the measured x-rays are emitted by rare earth elements. Sumbaev demonstrated that, for these elements, the energies of Kα transitions can differ by up to 0.5 eV between different compounds [34].
- x-ray energies could depend on how the atomic vacancy is created, even though this is not clearly demonstrated. The tabulated energies originate from atomic photo-ionisation and not nuclear decay [35].

Based on the method exposed in Section 4.1, the median values $\tilde{e}_{lin,i,j}$ have been calculated for the five calibration lines of each set $\epsilon_{i,j}$. Then, the non-linearity deviations, i.e. $\Delta\tilde{e}_{lin,i,j} = \tilde{e}_{lin,i,j} - e_{cal}$, are fitted by polynomial functions $f_{lin,i,j}$ (figure 9), where $e_{cal}$ is the calibration energy listed in table 1. The data points, $\Delta\tilde{e}_{lin,i,j}$, representing the deviations are weighted by the inverse of the quadratic sum of the variances:

$$\sigma^2_{\tilde{e}_{lin,i,j}} = \sigma^2_{stat,\tilde{e}_{lin,i,j}} + \sigma^2_{e_{cal}} + \sigma^2_{ROI} \quad \text{equation (6)}$$

where $\sigma_{e_{cal}}$ is the uncertainty of the tabulated energy (table 1).

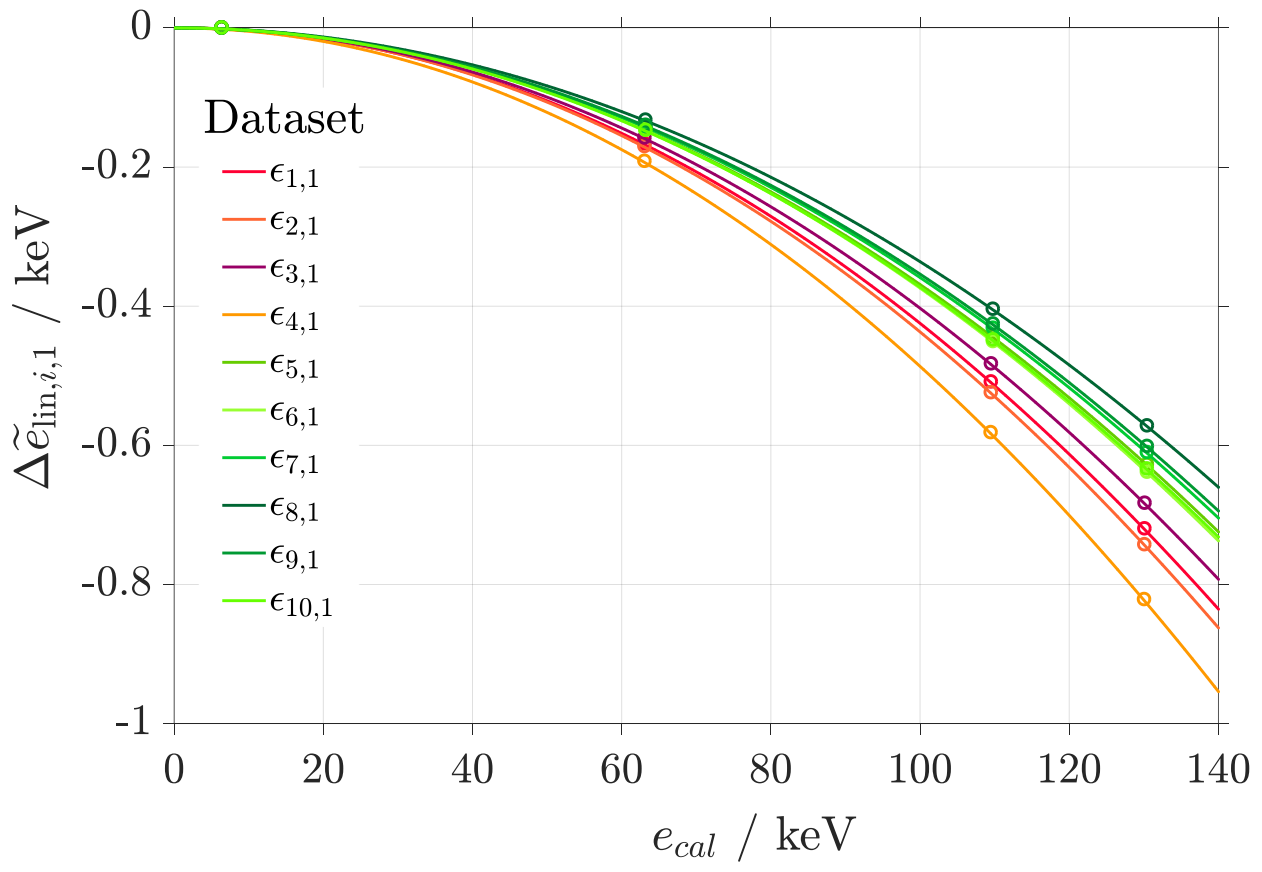

**Figure 9.** Second order polynomial functions $f_{poly,i,1}$ established for each dataset $\epsilon_{i,1}$ to fit the energy difference $\Delta\tilde{e}_{lin,i,1}$ during the measurement of the source 1. The datasets of the run R1 are from *i* = 1 to 4, red to yellow solid lines, and the datasets for run R2 from *i* = 5 to 10, green solid lines. The circles represent the experimental data; their uncertainties are much smaller than the circles and are not shown.

Figure 9 shows the $f_{poly,i,1}$ functions established for each $\epsilon_{i,1}$ dataset of the $S_1$ source, which was measured during the two sessions R1 and R2. The non-linearity is visually well described by a second-order polynomial. Each dataset has its own function that differs from the others due to variations in the thermodynamic properties of the MMC absorbers. Additionally, it can be seen that session R2 is more linear than R1 due to the lower sensitivity of the MMCs during this measurement (see Table 2).

These differences mean it is essential to correct each dataset individually using its own function. It was also decided that the datasets would be recalibrated after each change of source. A total of 40 calibration functions, $f_{poly,i,j}$ were established.

Figure 10 plots the residuals between the functions $f_{poly,i,j}$ and the energy differences, $\Delta\tilde{e}_{lin,i,j}$. For most datasets, the residuals are close to zero within the uncertainty range, which confirms the appropriateness of the second-order polynomial correction for non-linearity. Furthermore, 68% prediction intervals are calculated from these residuals. These intervals will be used to determine the uncertainty associated with the non-linearity correction. It can be seen that the intervals correctly cover the dispersion of the residuals, even though they are not compatible with zero.

The RMSE (root mean square error) of the residuals is also calculated to evaluate the quality of the fit, taking into account the number of degrees of freedom. In some cases, datasets with large RMSE can be explained. For the four datasets $\epsilon_{2,j}$, the slew rate of the SQUID electronics was insufficient to correctly measure large pulses with energies above 90 keV leading to poor non-linearity correction.

Some calibration points have residuals that are not consistent with zero, and this occurs more frequently with the 130 keV line. Possible reasons for these biases include:

- gain non-linearity in the SRS preamplifier between the SQUID electronics and the acquisition card. Certain preamplifiers exhibit greater non-linearity deviations when the lines correspond to signals at the edge of the preamplifier's range. This is likely the cause of the discrepancies observed in the residuals of the datasets $\epsilon_{4,j}$ and $\epsilon_{10,j}$, which share the same preamplifier showing a linearity issue.

- variations in detector sensitivity and gain, which are themselves caused by temperature fluctuations in the cryostat and small jumps in the gain. Since the highest energies are also the most sensitive to these variations, imperfect correction for these fluctuations leads to biases that increase with photon energy.

- the peak tails; these are themselves partly a consequence of the previous point, the tails increase with the photon energy,

- at a lesser extent, the integral non-linearity of the acquisition card.

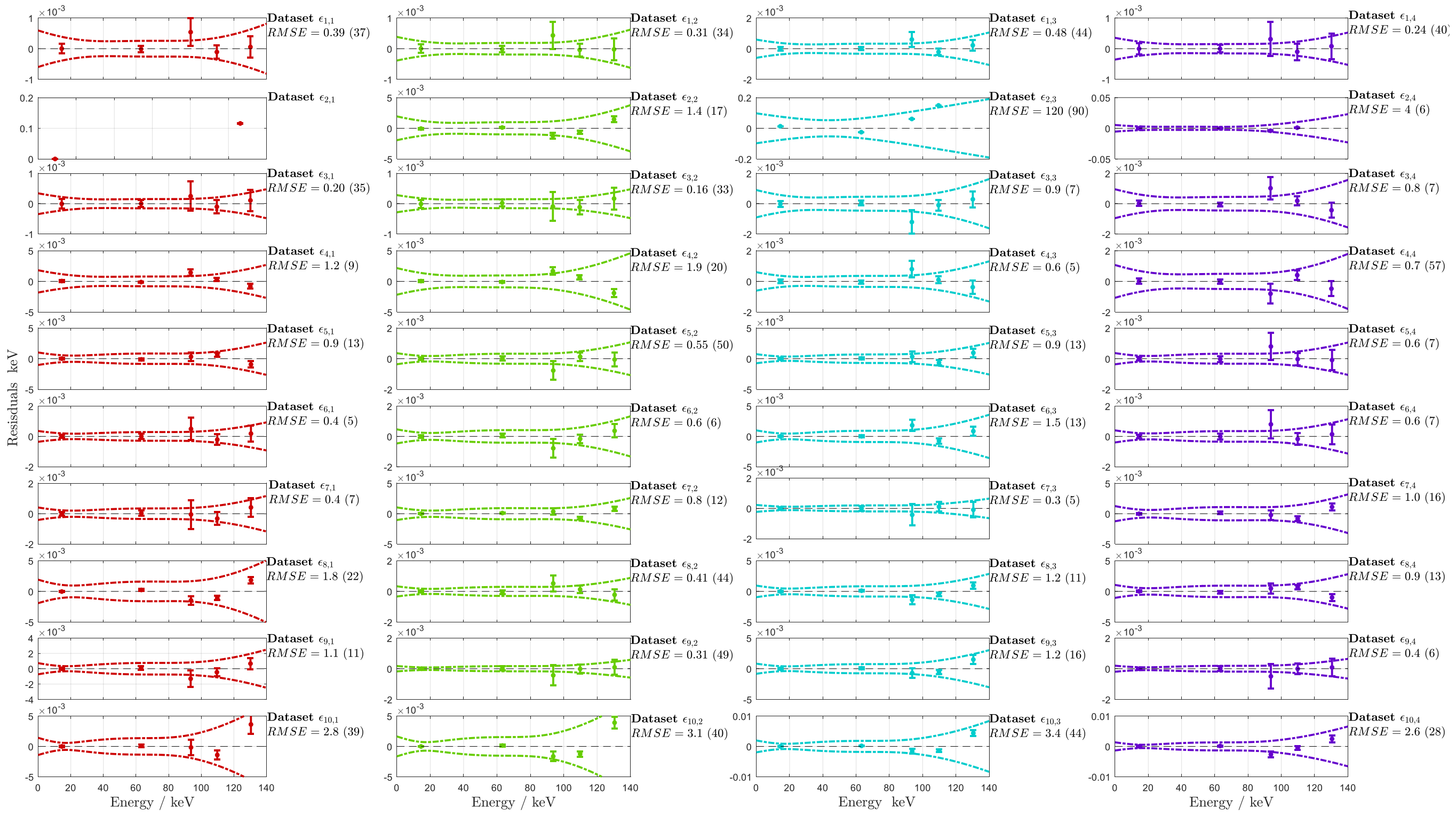

**Figure 10.** Residuals between the polynomial function $f_{poly,i,j}$ and the non-linearity deviation $\Delta\tilde{e}_{lin,i,j}$ obtained for the different datasets $\epsilon_{i,j}$. The dash-dotted lines are the 68% prediction intervals of the calibration polynomial functions. For each data set the root mean square error ($RMSE$) of the fit is given in eV. For the dataset $\epsilon_{2,1}$ the number of calibration points was insufficient to fit $\Delta\tilde{e}_{lin,i,j}$.

This non-linearity could be corrected by increasing the degree of the polynomial function; however, given the small number of calibration points, this reduces the degrees of freedom of the fit and could introduce systematic errors. Therefore, we opted to use a quadratic function by default, provided the prediction interval covers these residuals. Pixels with poor linearity will potentially be rejected by the analysis presented in the next section.

## 5 Method to assess the energies

### *5.1 Energies of the datasets and statistical analysis*

A statistical method was implemented to assess the energies by analysing the different datasets $\epsilon_{i,j}$. To determine the energy of each gamma-ray line in a dataset $\epsilon_{i,j}$, the same method as described in Section 4.1 was used. Specifically, events $e_{lin,i,j,k}$ are selected within a narrow ROI around the line with a cut-off parameter $C_{cut} = 2$ (see equation (4)). The energy is then the median $\tilde{e}_{lin,i,j}$ of the selected events composing the ROI reduced by $C_{cut}$. Finally, $\tilde{e}_{lin,i,j}$ is corrected for non-linearity using the function $f_{poly,i,j}$ and equation (7).

$$\tilde{e}_{poly,i,j} = \tilde{e}_{lin,i,j} - f_{poly,i,j}\left(\tilde{e}_{lin,i,j}\right) \qquad (7)$$

The associated variance $\sigma^2_{\tilde{e}_{poly,i,j}}$ is the quadratic sum of three components:

- the statistical uncertainty component of Type A, given by $\sigma_{stat,\tilde{e}_{i,j}}$= $\sigma_{stat,\tilde{e}_{lin,i,j}}$ and equation (5),
- the Type B uncertainty, $\sigma_{cal,\tilde{e}_{i,j}}$, is introduced by the non-linearity correction with the polynomial function $f_{poly,i,j}$. The standard deviation, $\sigma_{cal,\tilde{e}_{i,j}}$, corresponds to the prediction interval resulting from the residuals between $\Delta\tilde{e}_{lin}$ and the polynomial function $f_{poly,i,j}$, see figure (10),
- the Type B uncertainty, $\sigma_{ROI}$ introduced by the selection method.

For each gamma-ray line and a given source $S_j$, there are a maximum up to ten energies $\tilde{e}_{poly,i,j}$ from the corresponding ten datasets $\epsilon_{i,j}$. Figure 11 shows example of energies $\tilde{e}_{poly,i,j}$ for different gamma-ray lines, dataset $\epsilon_{2,j}$, and sources $S_j$. For some lines (i.e. 26. keV or 88.0 keV), $\tilde{e}_{poly,i,j}$ presents relatively large dispersion, that is in most cases covered by their associated uncertainties. In a few cases, some values $\tilde{e}_{poly,i}$ are clearly outliers; they have been excluded on the basis of a Z-score greater than 3. There are only few excluded data points, in average much less than one per gamma-ray line and dataset, because generally the data at the margin have also large prediction intervals for the non-linearity correction.

The data have also large differences in their respective uncertainties often largely dominated by the Type B uncertainty due to the non-linearity correction. Therefore, the data dispersion of $\tilde{e}_{poly,i}$ do not correspond to what is typically measured, i.e. experimental data with close Type A uncertainties, sharing the same Type B uncertainty. Here, the data dispersion is more similar to what is found in datasets of data evaluations or interlaboratory comparisons. Therefore, to estimate the energies and their uncertainties from each dataset, we applied statistics used for data evaluations or intercomparisons [36].This approach requires the assumption that the measurements for each pixel are independent which is not strictly guaranteed: the pixels share several measurement devices, the calibration sources and the analysis methodology. In particular, they are measured by the same models of MMC chip, SQUID electronics, preamplifiers and acquisition cards. These DAQ (data acquisition system) devices may share measurement characteristics and errors that will be correlated when we analyse the entire dataset. Since non-linearity correction is the main correction to our measurement, it is the most likely source of potential correlation and bias. However, when analysing the residuals of the calibration in Figure 10, the correlation between pixel residuals in the same column is not obvious, i.e. for the same source $S_j$ (datasets $\epsilon_{i,j}$ with $j$ constant and $i = 1, \ldots, 10$).

This is illustrated in the figures presented in the Appendix A, where each residual at a given energy is associated with a probability density following a normal distribution with a variance given by equation (6) and a mean value of $e_{cal} + \Delta\tilde{e}_{lin,i,j}$. The sum of these pixel probability densities is compared with that of the calibration line[4]. For datasets within a column in Figure 10, the maxima of the summed probability density functions of the ten pixels coincide with the maximum of the calibration line density. Any correlation in the errors would manifest as a shift between these maxima. Such an effect is observed in Appendix B when examining the sum of the probability densities of pixels across different sources (datasets $\epsilon_{i,j}$ with $i$ fixed and $j = 1, \ldots, 4$)), clear biases appear for pixels $\epsilon_{10,j}$ and, to a lesser extent, for pixel $\epsilon_{4,j}$. This is because they share the same electronic devices, which appear to exhibit a stronger non-linearity than those of the other pixels.

[4] It is worth noting that there are no significant differences between the envelope of the sum of the probability density functions and the sum of the probability density functions weighted by the uncertainty weights.

It can therefore be considered that the variability from the measurement conditions among different pixels is sufficient to treat the measurements as independent. Potential correlations between pixels are masked by variability among the: 1- electronic components (even if they are of the same model), 2- temperature correction methods in run R1 and R2, 3- sensor sensitivities (see Table 2), and 4- the baseline level of pulses.

The resulting assessed energy $\bar{E}_{w,j}$ of a transition is the result of a weighted average over the values $\tilde{e}_{poly,i}$ given by equation (8).

$$\bar{E}_{w,j} = \frac{\sum_i \tilde{e}_{poly,i,j} \times w_{i,j}}{\sum_i w_{i,j}} \qquad (8)$$

where $w_{i,j}$ is the weight of a dataset $i$ equal to $1/\sigma^2_{\tilde{e}_{poly,i,j}}$.

The Type A and the Type B uncertainties of $\bar{E}_{w,j}$, respectively $\sigma_{\bar{E}_{w,j,A}}$ and $\sigma_{\bar{E}_{w,j,B}}$, are calculated using the corresponding uncertainties of the non-excluded $k$ energies $\tilde{e}_{poly,i,j}$. The Type A uncertainty $\sigma_{\bar{E}_{w,j,A}}$ is a weighing of the statistical uncertainties $\sigma_{stat,\tilde{e}_{poly,k,j}}$ given by equations (9).

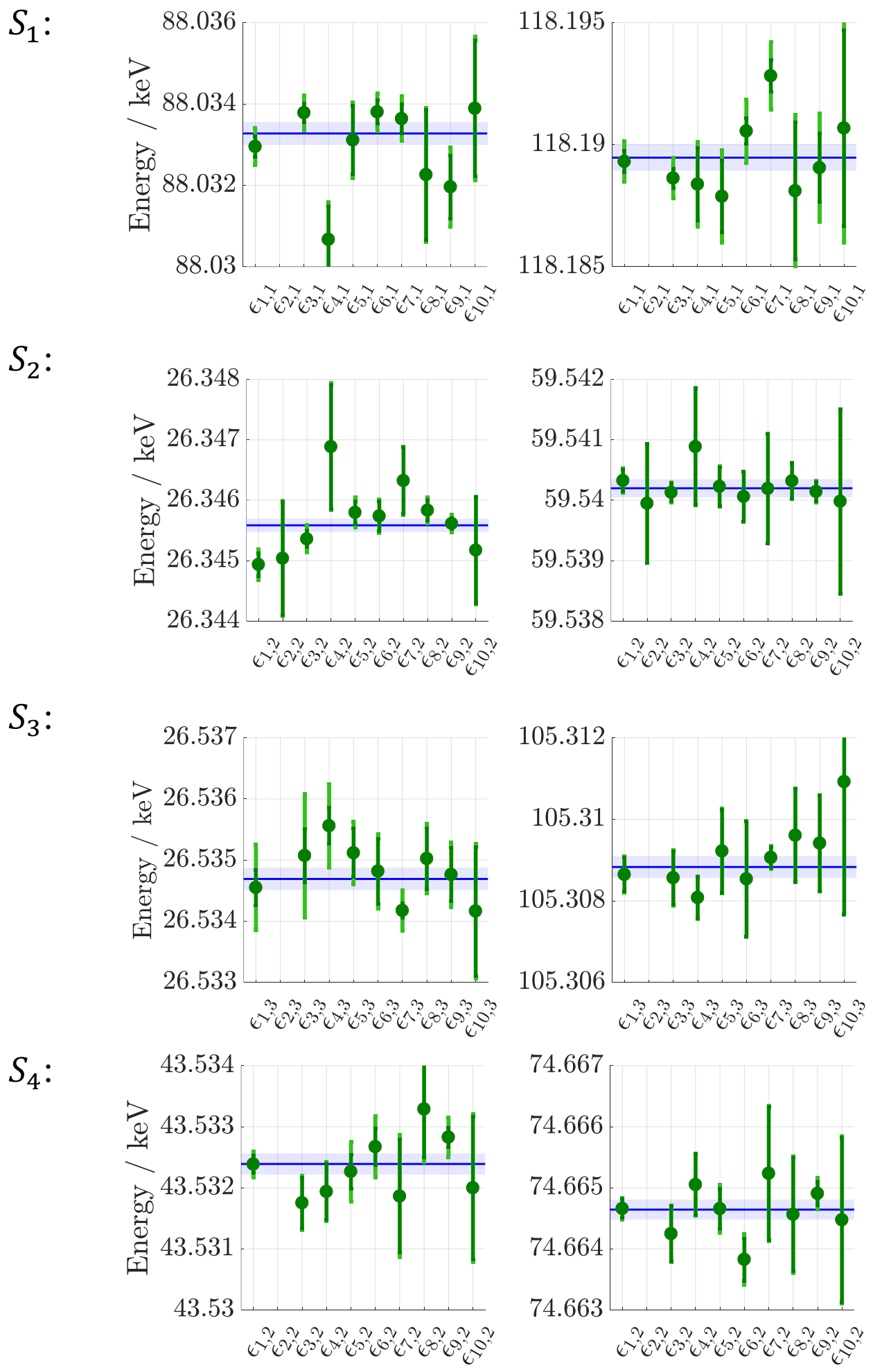

**Figure 11.** Example of energies $\tilde{e}_{poly,i}$ from the 10 datasets $\epsilon_{i,j}$ for two gamma line energies per source $S_j$; the light green error bars represent the total uncertainty with a coverage factor of $k$ = 1. The dark green error bars represent the Type B uncertainties. The blue line shows the reassessed energy $\bar{E}_{w,j}$ calculated using equation (8) and the energies $\tilde{e}_{poly,i}$. The blue band represents its uncertainty at $k$ = 1. There is no excluded outlier in these datasets.

$$\sigma_{A,\bar{E}_{w,j}} = \frac{1}{\sum_k w_{k,j}} \sqrt{\sum_k \left( w_{k,j}^2 \cdot \sigma_{stat,\tilde{e}_{poly,k,j}}^2 \right)} \qquad (9)$$

Type B uncertainty $\sigma_{\bar{E}_{w,j,B}}$, cannot be estimated by weighting the type B uncertainties $\sigma_{cal,\tilde{e}_{i,j}}$ and $\sigma_{ROI}$, because the measurement of $\tilde{e}_{poly,i,j}$ is based on the same method and some type B uncertainties are incompressible and common to all measurements, such as uncertainties on calibration lines $\sigma_{e_{cal}}$ or $\sigma_{ROI}$. Their contributions cannot be reduced by multiplying the number of pixels. Instead, the type B uncertainty, $\sigma_{B,\bar{E}_{w,j}}$, is evaluated according to equation (10), which provides a conservative estimate than equation (9).

$$\sigma_{B,\bar{E}_{w,j}} = \min \left( \left[ \sigma_{cal,\tilde{e}_{k,j}}^2 + \sigma_{ROI}^2 \right]^{\frac{1}{2}} \right) \qquad (10)$$

Figure 11 shows the reassessed energies $\bar{E}_{w,j}$ (blue line) and the associated uncertainties (blue band).

Since the counting statistics are similar among the pixels, the averaged value $\bar{E}_{w,j}$ is essentially weighted by Type B uncertainties. It is therefore necessary to ensure that these uncertainties are not underestimated due to hidden contributions. First, underestimated uncertainties lead to inconsistent datasets; this can be verified by comparing the so-called internal variance and external variance of the dataset [37]. After removing outliers, all datasets $\epsilon_{i,j}$ are found to be consistent.

Furthermore, an underestimation of Type B uncertainties, when they are dominant, may manifest as a discrepancy between the weighted average and the arithmetic average. Here, 77 % and 85 % of the arithmetic averages are in agreement with the weighted averages within their uncertainties for $k$=1 and $k$=2, respectively. Most of the discrepant data (8%) concern the 136 keV lines, which are at the limit of the calibrated energy range. There is good agreement between these two averages confirming that the Type B uncertainties are correctly estimated and supporting the assumption that the data can be considered independent.

Other evaluation approaches were explored based on the Paul-Mandel statistical procedure [38,39]. This procedure consists of checking the consistency of a dataset; if the dataset is found to be inconsistent, the uncertainties are increased iteratively until consistency is achieved. This procedure yielded results compatible with $\bar{E}_{w,j}$ (equation (8)) within uncertainties, and the associated uncertainties are comparable to, or even smaller than, those obtained using equations (9) and (10).

### *5.2 Validation of the method*

The consistency of the assessed energies, $\bar{E}_{w,j}$, obtained using the method detailed in Section 5.1, was verified in different ways:

1- comparing the measured energies $\bar{E}_{w,j}$ with some recommended values that have been measured by WDS but are not currently used for calibration,

2- comparing the gamma line energies, $\bar{E}_{w,j}$, of radionuclides present in different sources $S_j$,

3- comparing the gamma line energies $\bar{E}_{w,j}$, with different gamma-transition cascade paths, where the decay scheme allows.

Figure 12 compares the assessed energies $\bar{E}_{w,j}$ obtained from different sources $S_j$ with the recommended values based on precise WDS data. These energies relate to lines other than those used for calibration[5], enabling an independent comparison. Almost all of the $\bar{E}_{w,j}$ energies are in agreement with the recommended values at $k$ = 1.

Furthermore, the $\bar{E}_{w,j}$ energies are consistent with each other at $k$ = 1, meaning that the results are reproducible from one source to another. Regarding the transition γ1,0 of the $^{57}$Co decay, the gamma-ray line cannot be re-assessed directly because it is a calibration line. However, it can be obtained independently by subtracting the energies of the lines γ2,0 and γ2,1. The first panel of Figure 12 shows this difference obtained for the four sources measured. For three of the sources, the values agree at $k$ = 1 with the recommended value and with the calibration value of 14.412492 (19) keV (all values agree at $k$ = 2), whereas the recommended value itself does not agree with the calibration value.

[5] In the case of the 14.4 keV energy, the measured energy is deduced by the difference between the 136.5 keV and the 122.1 keV lines, independently of the calibration line at 14.4 keV.

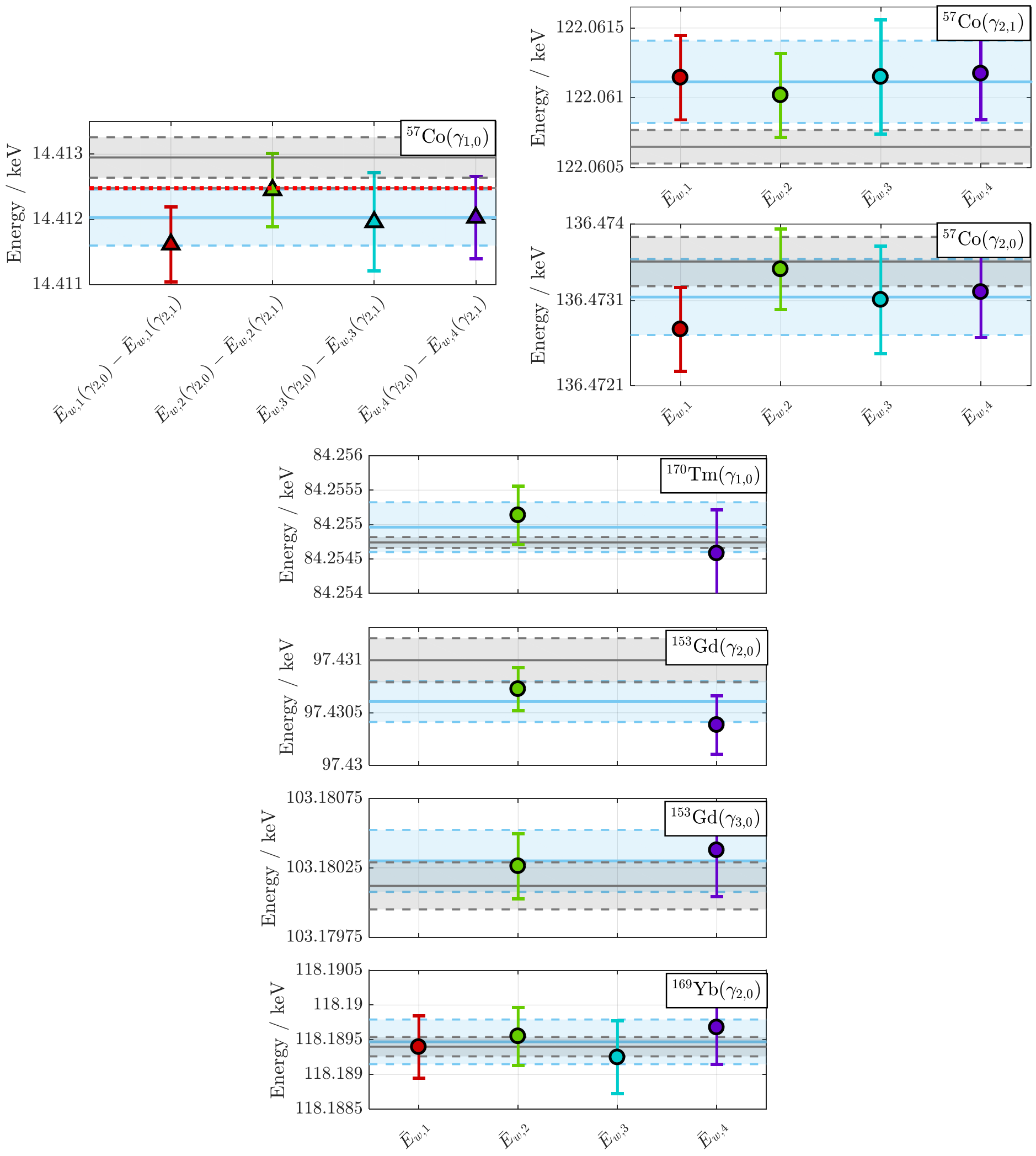

**Figure 12.** Reassessed energies $\bar{E}_{w,j}$ for different sources $S_j$ (coloured data points). The grey lines are the recommended energies and the dashed grey lines represent their uncertainties [17]. The blue lines represent the final results $E_f$ in Table 3, and the dashed blue lines represent their uncertainties. Uncertainties are given with a coverage factor of $k$ = 1. The dotted red line shows the value used for the calibration at 14.412492 (19) keV, and its uncertainty is smaller than the line width.

Several other measured gamma-ray energies can be verified through gamma cascades between different gamma-excited levels. This is the case for transitions in $^{155}$Eu, $^{241}$Am and $^{133}$Ba. Figure 13 shows the case of three gamma transitions of $^{237}$Np following the decay of $^{241}$Am, involving γ1,0, γ2,0 and γ2,1. A good agreement is achieved for the three energies between the different cascade combinations. The difference between γ2,0 and γ2,1 results in a lower uncertainty for γ1,0 than the direct measurement of γ1,0.

The gamma transition cascades of $^{155}$Gd in the decay $^{155}$Eu are particularly useful for checking the detector linearity, as the transitions cover a broad energy range from 18 keV to 105 keV with well-spaced gamma-ray energies. The cascades involve three excited levels and six gamma-ray energies to reach the ground state. Good agreement is verified for all the energies and for the different cascade combinations.

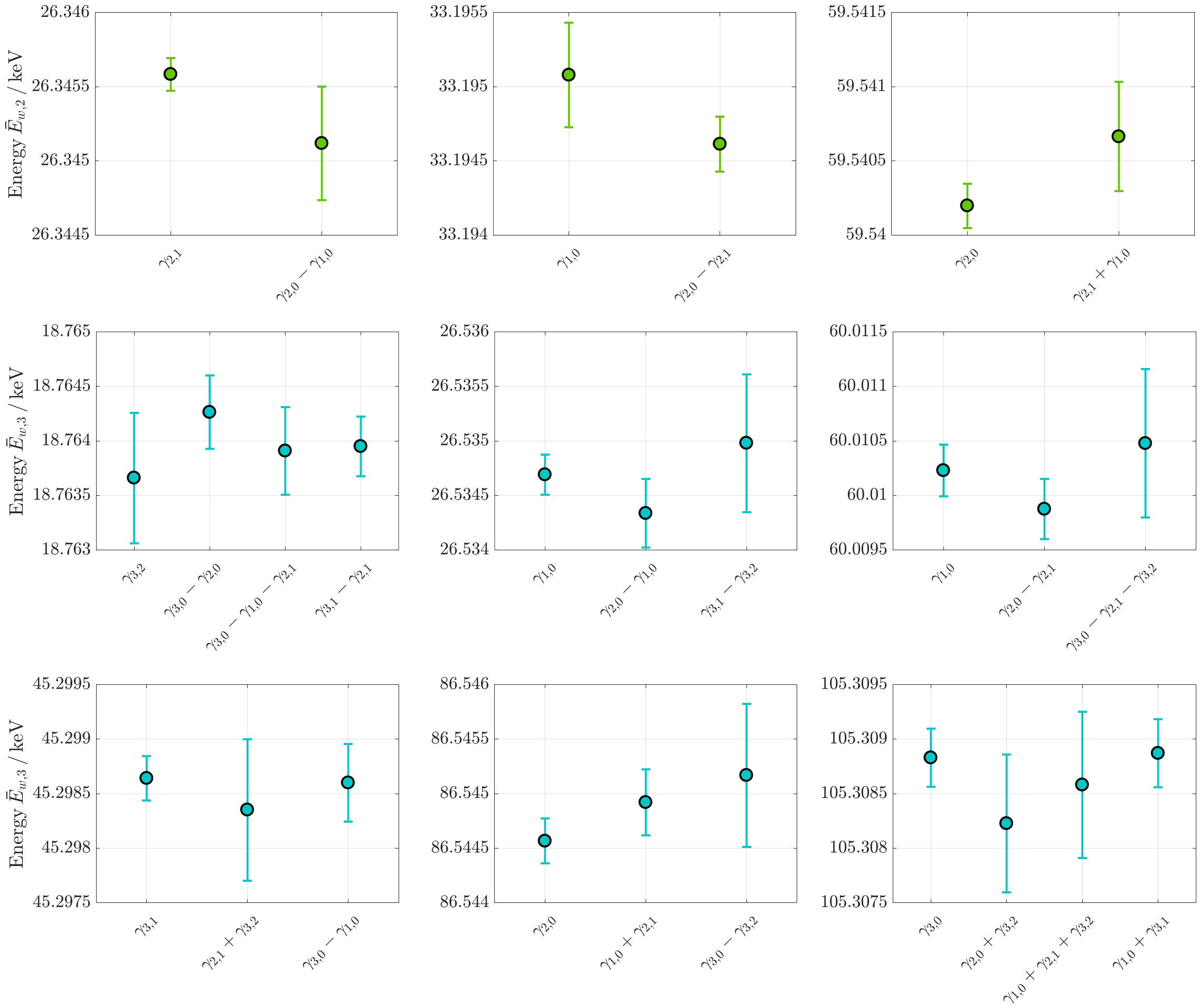

**Figure 13.** Reassessed energies $\bar{E}_{w,j}$ of gamma-ray lines compared with sums or differences of gamma transitions for $^{241}$Am (green) and $^{155}$Eu (blue). The parity and the nomenclatures of the transitions are detailed in Table 3. Uncertainties are given with a coverage factor of $k$ = 1.

## 6 Results

For a given transition, the final energies $E_f$ provided in this work are equal to the assessed energies obtained in the previous section when only one value, $\bar{E}_{w,j}$, has been measured. When several energies $\bar{E}_{w,j}$ are available for a transition - either because the radionuclide was present in multiple sources $S_j$, or because the gamma energy can be deduced from various gamma-excited levels - the final value $E_f$ is the weighted average of the measured values $\bar{E}_{w,j}$. If the set includes values $\bar{E}_{w,j}$ from different sources, then the type B uncertainty cannot be lower than the minimum value of the type B uncertainty of the $\bar{E}_{w,j}$ values because these uncertainties may be correlated. The results are given in Table 3, which also shows the energies $\bar{E}_{w,j}$ used to compute $E_f$ and the level combinations when they are derived through a gamma cascade.

Figure 12 compares the final energies $E_f$ (blue lines) for some selected transitions with the recommended ones, which were established using data from wavelength dispersive spectrometry and evaluated by Helmer and van der Leun [17]. As discussed in Section 5.1 regarding the energies $\bar{E}_{w,j}$, all of the present measured energies $E_f$ are consistent with the recommended values, except for the γ2,1 and the γ1,0 transitions of the $^{57}$Co decay, where a slight disagreement is observed at $k$ = 1, though this remains compatible at $k$ = 2.

Table 3 summarises the results of the gamma-ray energies $E_f$ compared with the recommended values $E_{rec}$. Most of the recommended values $E_{rec}$ are identical to those recommended by the DDEP (Decay Data Evaluation Project, [23]) and the ENSDF (Evaluated Nuclear Structure Data File, [40]), as well as those from Helmer and van der Leun [17], except for the gamma rays of $^{155}$Eu [41]. The column "Deviation ($E_f$ - $E_{rec}$)" and its associated uncertainty allow us to judge the agreement between the values: when the uncertainty is smaller than the deviation, the result of the present work is not compatible with the recommended value at a coverage factor of $k = 1$. While our data are in good agreement with the recommended values established using WDS data ($^{57}$Co, $^{153}$Gd, $^{169}$Yb and $^{170}$Tm), there is clear disagreement with some values established using EDS data. These include $^{241}$Am (gamma rays of $^{237}$Np) and $^{155}$Eu (gamma rays of $^{155}$Gd) and $^{239}$Np (gamma rays of $^{239}$Pu).

For $^{241}$Am, the recommended gamma-ray energies of 26.3 keV and 59.5 keV are particularly accurate for EDS measurements using semiconductors. The 33 keV value is derived from the difference between these two gamma-ray energies. These values are based on measurements performed by Helmer in 1993 which disagreed with the recommended values at the time of 59.5364 (10) keV [42]. In 2020, G. B. Kim *et al*. measured this gamma-ray line at 59.5393 (4) keV using high energy resolution MMCs [22], which is clearly incompatible with that of Helmer, but is almost consistent with our own measurement of 59.54026 (14) keV at k = 2.

Regarding the $^{155}$Eu, the recommended energies from ENSDF and DDEP differ. Our measurements tend to validate those from ENSDF, with which better agreement is found. This is particularly evident for the gamma ray γ2,0 at 86.5 keV, where significant discrepancies are observed with the DDEP value. Interestingly, the 86.5 keV has intense escape peaks from the K$\alpha$ x-ray escapes of the gold absorber in our spectra, and the escape peak energies confirm our full energy peak value [18].

For $^{133}$Ba, the γ2,0 transition at 160.6 keV can be measured directly, as well as through the summed cascade of the γ2,1 gamma-ray transition and the γ1,0 gamma-ray transition. Despite the low statistics in the spectrum representing a major contribution to the uncertainty, there is good agreement between the two values (see Table 3).

For $^{239}$Np, there is a significant discrepancy between the current result and the recommended value for the gamma-ray line γ4,2. The latter is likely constrained by a WDS measurement with an uncertainty of 2 eV [43]. However, few details are available regarding the uncertainty budget of this measurement.

The uncertainties of the final energies $E_f$ are in the order of 0.2 eV, which greatly improves upon the energies previously determined by EDS with semiconductor detectors, except for $^{241}$Am. In this case, however, there is a clear discrepancy between the different measurements, namely those from Helmer and Kim, and the present result [22,42]. For $^{243}$Am, the uncertainties on two measured gamma-ray energies have been improved by two orders of magnitude.

As shown in Table 3, our results are frequently dominated by the Type B uncertainty associated with the non-linearity correction using the polynomial function $f_{poly,i,j}$. This uncertainty is a combination of the calibration energy uncertainty, $\sigma_{e_{cal}}$, the uncertainty in determining the energy in a ROI, $\sigma_{ROI}$, and the statistical uncertainty in the calibration line ROI, $\sigma_{stat}$. However, the Type B variance of the non-linearity correction is larger than the quadratic sum of $\sigma_{e_{cal}}$ and $\sigma_{ROI}$ and $\sigma_{stat}$.

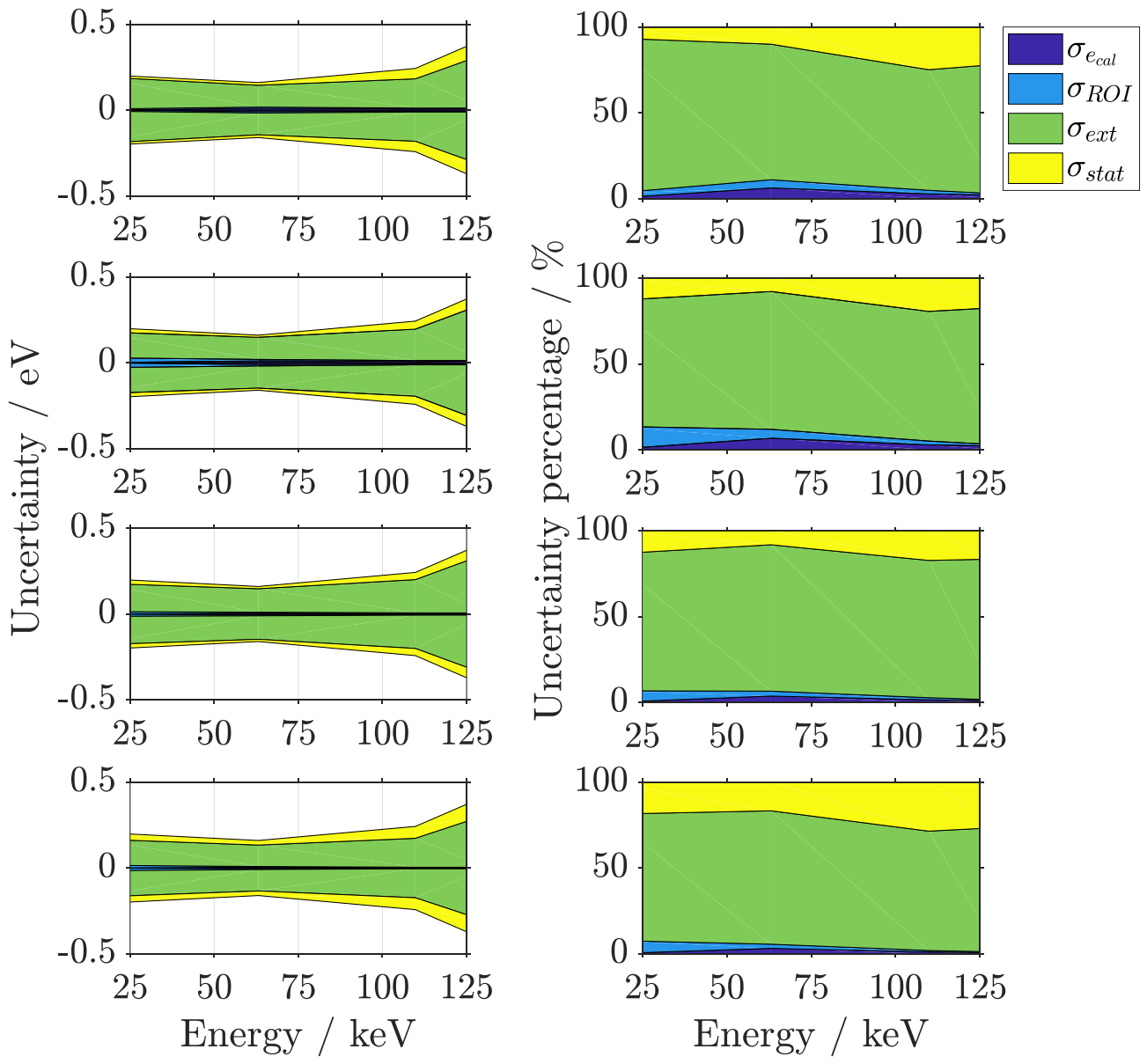

**Figure 14.** Absolute and relative uncertainty contributions from the non-linearity corrections; $\sigma_{e_{cal}}$ is the uncertainty of the calibration lines, $\sigma_{ROI}$ is the uncertainty due to the analysis method, and on the calibration lines, $\sigma_{stat}$ is the statistical uncertainty and $\sigma_{ext}$ is the uncertainty from unidentified extrinsic contributions.

As discussed in Section 4.2, another uncertainty component $\sigma_{ext}$ originates from multiple sources of non-linearity, such as:

- the pre-amplifier, which is probably the case in figure 10, for the datasets $\epsilon_{10,j}$,

- the integral non-linearity of the analogue-to-digital conversion of the acquisition card,
- the temperature drift correction when the temperature change cannot be considered small compared to the base temperature, or when the drift is over or under corrected.

Figure 14 shows the absolute and relative contributions of the different uncertainty components. The uncertainty $\sigma_{ext}$ is clearly the dominant one over the entire energy range. Based on this observation, the results presented can be improved so that the statistical and the calibration energy uncertainties become the dominant factors in future experiments. The non-linearity from the DAQ could be reduced by precisely calibrating the preamplifiers and by using an 18-bit data acquisition card. In addition, the peak tails can be reduced by mitigating the thermal crosstalk within the MMC chips and by achieving better control over sensitivity drift due to temperature fluctuations. This last point is probably the most difficult to address.

## 7 Conclusions

A high-energy-resolution magnetic microcalorimeter spectrometer has been developed to accurately measure gamma-transition energies below 150 keV. The spectrometer has eight pixels and achieves a FWHM energy resolution ranging from 15 eV to 30 eV. This allows negligible statistical uncertainties to be reached for the measured gamma-ray lines.

The non-linearity of the detectors was corrected using five well-known gamma-ray lines emitted by $^{57}$Co and $^{169}$Yb between 14 keV and 130 keV. This was achieved using a second order polynomial function.

New experimental gamma-ray energies have been obtained for 27 transitions and 12 radionuclides ($^{57}$Co, $^{109}$Cd, $^{133}$Ba, $^{153}$Gd, $^{154}$Eu, $^{155}$Eu, $^{169}$Yb, $^{170}$Tm, $^{210}$Pb, $^{239}$Np, $^{241}$Am and $^{243}$Am). For most of these, the uncertainty is in the order of 0.2 eV, corresponding at best to a relative uncertainty of 1 ppm. This uncertainty is primarily due to the non-linearity correction and some extrinsic uncertainties, possibly arising from non-linearities in the DAQ and electronics. This could be improved in the future through measurements carried out with a more linear and better-characterised acquisition system.

The energies measured in the present study are in good agreement with those measured by wavelength dispersion spectrometry, which is renowned for its accuracy, linearity and link to an absolute energy scale. In addition, the present measurements were validated by comparing gamma transition energies at different energy levels.

These measurements greatly improve the accuracy of around 20 gamma-ray energies that were previously determined using semiconductor detectors. These energies relate to radionuclides with relatively long half-lives, which are difficult to measure using wavelength-dispersive spectrometry due to their low detection efficiency. Therefore, we have demonstrated the capabilities of low-temperature detectors in measuring the gamma-ray energies of radionuclides precisely up to 140 keV, which can be used for the energy calibration of other spectrometers, especially in the development of different fields of fundamental physics. These new energy values are now available for data evaluation to improve uncertainties in the tables.

**Table 3.** Results of the energies obtained in this work compared with the recommended energies $\boldsymbol{E_{rec}}$ for the different measured radionuclides[6]. The Type A and Type B uncertainties are detailed, as are the energies $\boldsymbol{\bar{E}_{w,j}}$ used to compute $\boldsymbol{E_{rec}}$. The recommended values are from the DDEP, except for $^{155}$Eu where both DDEP and ENSDF recommended values are presented, The superscript symbol * indicates data are from ENSDF. The superscript symbol † indicates recommended values evaluated with data measured using WDS.

| Parent | Daughter | Initial level Jiπ | Final level Jfπ | Levels $\gamma i,f$ [23] | Recommended energy, $E_{rec}$ / keV | Energy of this work, $E_f$ / keV | Relative uncertainty / ppm | Percentage contribution of Type A / % | Percentage contribution of Type B / % | Deviation $E_f$ - $E_{rec}$ /eV | Reassessed energies, $\bar{E}_{w,j}$ / keV[7] | Level combinations $\gamma i,f$ | Source $S_j$ |
|---|---|---|---|---|---|---|---|---|---|---|---|---|---|
| $^{57}$Co | $^{57}$Fe | 3/2- | 1/2- | γ1,0 | 14.41294 (31)† | 14.4120 (5) | 48.3 | 4.7 | 95.3 | -0.9 (6) | 14.4116 (8) | 2,1 - 2,0 | *S1* |
| | | | | | | | | | | | 14.4125 (8) | 2,1 - 2,0 | *S2* |
| | | | | | | | | | | | 14.4120 (10) | 2,1 - 2,0 | *S3* |
| | | | | | | | | | | | 14.4120 (8) | 2,1 - 2,0 | *S4* |
| | | 5/2- | 3/2- | γ2,1 | 122.06061 (12)† | 122.06107(30) | 2.4 | 1.3 | 98.7 | 0.46 (32) | 122.06110 (30) | 2,1 | *S1* |
| | | | | | | | | | | | 122.06098 (30) | 2,1 | *S2* |
| | | | | | | | | | | | 122.06111 (41) | 2,1 | *S3* |
| | | | | | | | | | | | 122.06113 (33) | 2,1 | *S4* |
| | | 5/2- | 1/2- | γ2,0 | 136.47351 (29)† | 136.47309 (45) | 3.3 | 6.1 | 93.9 | -0.4 (5) | 136.47271 (49) | 2,0 | *S1* |
| | | | | | | | | | | | 136.47342 (47) | 2,0 | *S2* |
| | | | | | | | | | | | 136.4731 (6) | 2,0 | *S3* |
| | | | | | | | | | | | 136.4732 (5) | 2,0 | *S4* |
| $^{109}$Cd | $^{109}$Ag | 7/2+ | 1/2- | γ1,0 | 88.0336 (10) | 88.03321 (22) | 2.5 | 51.6 | 48.4 | -0.4 (10) | 88.03321 (22) | 1,0 | *S1* |
| $^{133}$Ba | $^{133}$Cs | 5/2+ | 5/2+ | γ2,1 | 79.6142 (19) | 79.61723 (37) | 4.7 | 69.3 | 30.7 | 3.1 (19) | 79.61723 (37) | 2,1 | *S4* |
| | | 5/2+ | 7/2+ | γ1,0 | 80.9979 (11) | 80.99649 (22) | 2.7 | 11.9 | 88.1 | -1.4 (11) | 80.99649 (22) | 1,0 | *S4* |
| | | 5/2+ | 7/2+ | γ2,0 | 160.6120 (16) | 160.6136 (5) | 3.1 | 39.4 | 60.6 | 1.5 (17) | 160.6117 (20) | 2,0 | *S4* |
| | | | | | | | | | | | 160.6137 (5) | 2,1 + 1,0 | *S4* |
| $^{153}$Gd | $^{153}$Eu | 5/2- | 5/2+ | γ2,0 | 97.43096 (21)† | 97.43057 (20) | 2.0 | 30.2 | 69.8 | -0.39 (29) | 97.43069 (20) | 2,0 | *S2* |
| | | | | | | | | | | | 97.43035 (28) | 2,0 | *S4* |
| | | 3/2+ | 5/2+ | γ3,0 | 103.18008 (17)† | 103.18026 (23) | 2.2 | 35.5 | 64.5 | 0.18 (28) | 103.18022 (24) | 3,0 | *S2* |
| | | | | | | | | | | | 103.18034 (33) | 3,0 | *S4* |

[6] According to the CODATA constants, we used the most recent conversion factor of 1.239841984×10$^{-6}$ eV.m for wavelength units to energy units in our results, as well as the recommended values. The recommended values of Helmer and van der Leun used a conversion factor of 1.23984244×10$^{-6}$ eV·m, they were updated in this table according to the recent conversion factor [44].

[7] When a gamma-ray energy is calculated through different gamma-ray transitions, the recoil energy, even small at these energies, was taken into account.

| | | | | | | | | | | | | | |
|---|---|---|---|---|---|---|---|---|---|---|---|---|---|
| $^{154}$Eu | $^{154}$Sm | 2+ | 0+ | γ1,0 | 123.0706 (9) | 123.0726 (7) | 6.0 | 68.3 | 31.7 | 2.1 (12) | 123.0726 (7) | 1,0 | *S3* |
| $^{155}$Eu | $^{155}$Gd | 2/3+ | 5/2+ | γ3,2 | 18.7630 (20)<br>18.7630 (20)* | 18.76397 (22) | 11.9 | 38.0 | 62.0 | 1.0 (20)<br>1.0 (20)* | 18.7637 (6) | 3,2 | *S3* |
| | | | | | | | | | | | 18.76395 (34) | 3,1 - 2,1 | *S3* |
| | | | | | | | | | | | 18.7645 (6) | 3,0 - 2,0 | *S3* |
| | | | | | | | | | | | 18.76393 (40) | 3,0 - 1,0 - 2,1 | *S3* |
| | | 5/2+ | 5/2- | γ2,1 | 26.531 (21)<br>26.527 (21)* | 26.53464 (17) | 6.3 | 60.3 | 39.7 | 4 (21)<br>8 (21)* | 26.53469 (19) | 2,1 | *S3* |
| | | | | | | | | | | | 26.5341 (5) | 2,0 - 1,0 | *S3* |
| | | | | | | | | | | | 26.5350 (7) | 3,1 - 3,2 | *S3* |
| | | 3/2+ | 5/2- | γ3,1 | 45.2990 (10)<br>45.2989 (10)* | 45.29860 (17) | 3.8 | 28.9 | 71.1 | -0.4(10)<br>-0.3(10)* | 45.29864 (20) | 3,1 | *S3* |
| | | | | | | | | | | | 45.29862 (36) | 3,0 - 1,0 | *S3* |
| | | | | | | | | | | | 45.2983 (6) | 3,2 + 2,1 | *S3* |
| | | 5/2- | 3/2- | γ1,0 | 60.0086 (10)<br>60.0089 (10)* | 60.01022 (19) | 3.1 | 29.4 | 70.6 | 1.6 (10)<br>1.3 (10)* | 60.01023 (24) | 1,0 | *S3* |
| | | | | | | | | | | | 60.01021 (33) | 3,0 - 1,0 | *S3* |
| | | | | | | | | | | | 60.0105 (7) | 3,0 - 3,2 -2,1 | *S3* |
| | | 5/2+ | 3/2- | γ2,0 | 86.5479 (10)<br>86.5463 (10)* | 86.54458 (16) | 1.9 | 25.6 | 74.4 | 3.3 (10)<br>-1.7 (10)* | 86.54438 (25) | 2,0 | *S2* |
| | | | | | | | | | | | 86.54457 (20) | 2,0 | *S3* |
| | | | | | | | | | | | 86.54491 (34) | 2,1 + 1,0 | *S3* |
| | | | | | | | | | | | 86.5452 (7) | 3,0 - 3,2 | *S3* |
| | | | | | | | | | | | 86.54436 (44) | 2,0 | *S4* |
| | | 3/2+ | 3/2- | γ3,0 | 105.3083 (10)<br>105.3087 (10)* | 105.30880 (13) | 1.3 | 50.9 | 49.1 | 0.5 (10)<br>0.1 (10)* | 105.3092 (7) | 3,0 | *S2* |
| | | | | | | | | | | | 105.30883 (27) | 3,0 | *S3* |
| | | | | | | | | | | | 105.30885 (31) | 3,1 + 1,0 | *S3* |
| | | | | | | | | | | | 105.3086 (7) | 3,2 + 2,1 + 1,0 | *S3* |
| | | | | | | | | | | | 105.3082 (6) | 2,0 + 3,2 | *S3* |
| $^{169}$Yb | $^{169}$Tm | 9/2- | 7/2- | γ11,8 | 93.61444 (8)† | 93.61436 (32) | 3.4 | 56.7 | 43.3 | -0.07 (33) | 93.61436 (32) | 11,8 | *S3* |
| | | 5/2+ | 1/2+ | γ2,0 | 118.18936 (14)† | 118.18941 (23) | 2.0 | 74.5 | 25.5 | 0.06 (27) | 118.18919 (39) | 2,0 | *S1* |
| | | | | | | | | | | | 118.18965 (33) | 2,0 | *S2* |
| | | | | | | | | | | | 118.1892 (5) | 2,0 | *S3* |
| | | | | | | | | | | | 118.1895 (9) | 2,0 | *S4* |
| $^{170}$Tm | $^{170}$Er | 2+ | 0+ | γ1,0 | 84.25471 (8)† | 84.25493 (36) | 4.3 | 83.1 | 16.9 | 0.22 (37) | 84.25510 (42) | 1,0 | *S2* |
| | | | | | | | | | | | 84.2545 (6) | 1,0 | *S4* |
| $^{210}$Pb | $^{210}$Bi | 0- | 1- | γ1,0 | 46.5390 (10) | 46.53818 (15) | 3.2 | 15.3 | 84.7 | -0.8 (10) | 46.53818 (15) | 1,0 | *S2* |
| $^{239}$Np | $^{239}$Pu | 7/2- | 7/2+ | γ8,6 | 61.4600 (20)† | 61.45635 (36) | 5.8 | 82.2 | 17.8 | -3.6 (20) | 61.45635 (36) | 8,6 | *S4* |
| | | 7/2- | 5/2+ | γ4,2 | 106.1250 (20)† | 106.11781 (25) | 2.3 | 25.0 | 75.0 | -7.2 (20) | 106.11781 (25) | 4,2 | *S4* |
| $^{241}$Am | $^{237}$Np | 5/2- | 7/2+ | γ2,1 | 26.34459 (20) | 26.34555 (11) | 4.0 | 29.3 | 70.7 | 0.95 (23) | 26.34557 (11) | 2,1 | *S2* |
| | | | | | | | | | | | 26.34511 (42) | 2,0 - 1,0 | *S2* |
| | | 7/2+ | 5/2+ | γ1,0 | 33.19629 (30) | 33.19477 (21) | 6.4 | 36.7 | 63.3 | -1.52 (37) | 33.19507 (35) | 1,0 | *S2* |
| | | | | | | | | | | | 33.19460 (27) | 2,0 - 2,1 | *S2* |
| | | 5/2- | 5/2+ | γ2,0 | 59.54088 (10) | 59.54023 (14) | 2.4 | 13.0 | 87.0 | -0.64 (17) | 59.54018 (15) | 2,0 | *S2* |
| | | | | | | | | | | | 59.54064 (40) | 2,1 + 1,0 | *S2* |
| $^{243}$Am | $^{239}$Np | 5/2- | 7/2+ | γ3,1 | 43.530 (20) | 43.53221 (18) | 4.2 | 37.0 | 63.0 | 2 (20) | 43.53221 (18) | 3,1 | *S4* |
| | | 5/2- | 5/2+ | γ3,0 | 74.660 (20) | 74.66461 (16) | 2.2 | 12.5 | 87.5 | 5 (20) | 74.66461 (16) | 3,0 | *S4* |

## Appendix A

For each source $S_j$, the residuals of the pixels $i$ from the datasets $\epsilon_{i,j}$ in Figure 10 are shown in panels (a) as probability densities following a Gaussian distribution, centred at $e_{cal} + \Delta\tilde{e}_{lin,i,j}$ and with a variance $\sigma^2_{\tilde{e}_{lin,i,j}}$(Equation 6). The amplitude of the Gaussian for each pixel $i$ is weighted by the inverse of its variance. The second panels (b) for a given source compares the sum of the pixel probability densities (in red) with that of the calibration line (in green), centred at $e_{cal}$ and with variance $\sigma^2_{e_{cal}}$. Good agreement is observed between the maxima of these two probability densities for the different energies and sources.

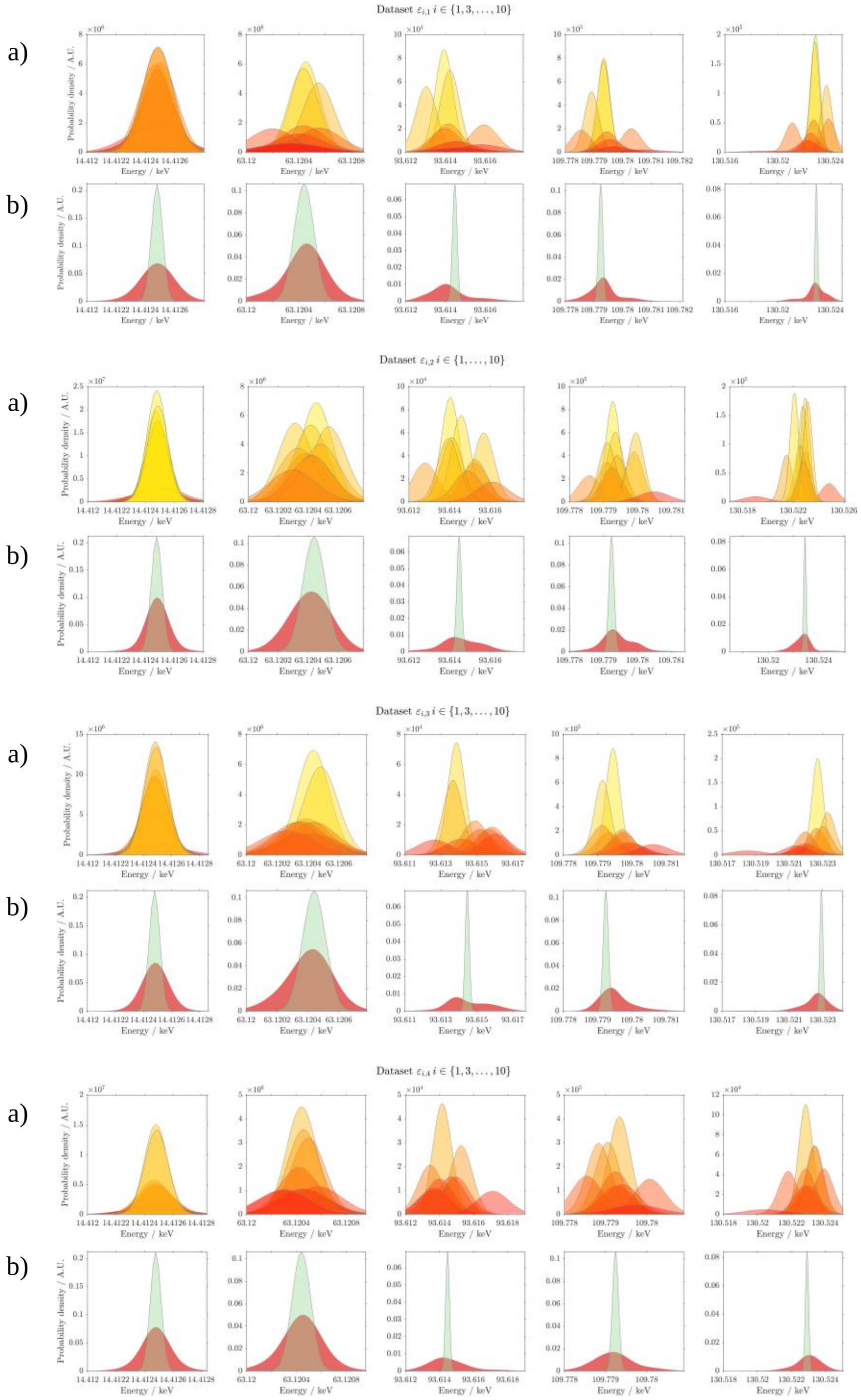

## Appendix B

For each pixel $i$, the residuals of the different sources $S_j$ from the datasets $\epsilon_{i,j}$ in Figure 10 are summed as probability densities (red) following a Gaussian distribution, centred at $e_{cal} + \Delta\tilde{e}_{lin,i,j}$ and with a variance $\sigma^2_{\tilde{e}_{lin,i,j}}$(Equation 6). The amplitude of the Gaussian for each pixel $i$ is weighted by the inverse of its variance. The sum of the pixel probability densities (in red) is compared with that of the calibration line (in green), centred at $e_{cal}$ and with variance $\sigma^2_{e_{cal}}$. For certain energies and pixels, clear discrepancies between the maxima of these two probability densities indicate the presence of a measurement bias.

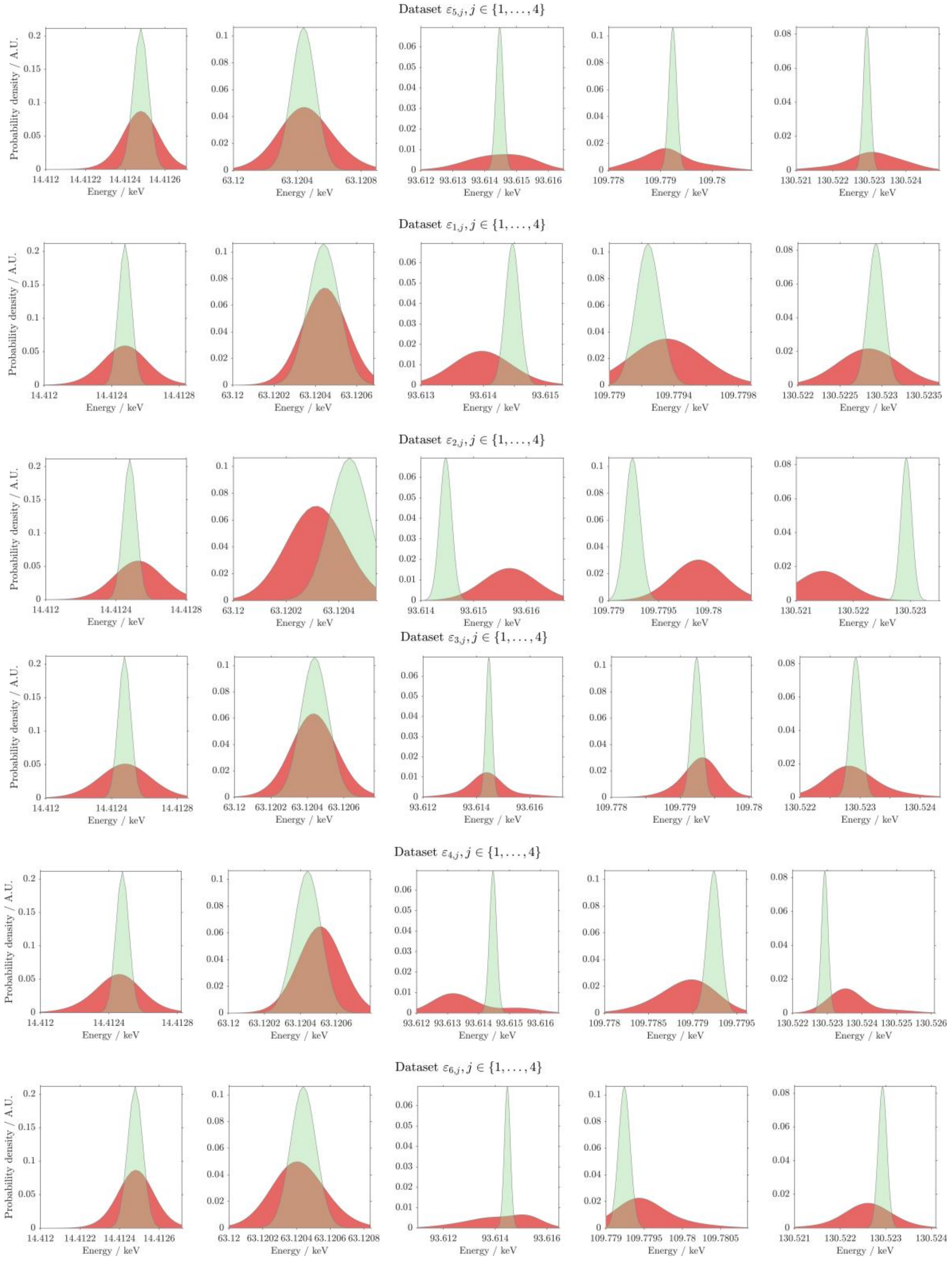

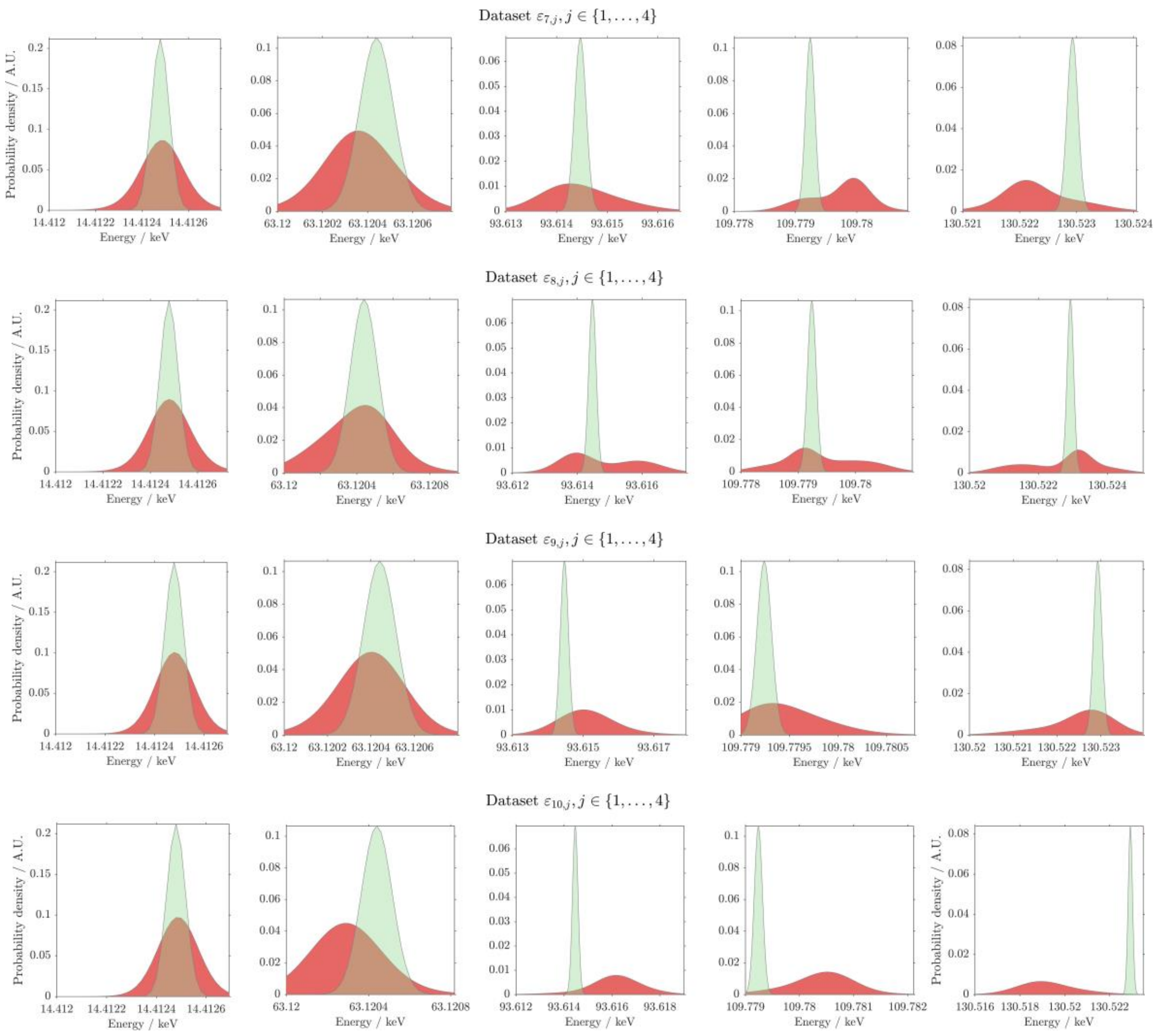
Dataset $\varepsilon_{7,j}, j \in \{1,\dots,4\}$
Dataset $\varepsilon_{8,j}, j \in \{1,\dots,4\}$
Dataset $\varepsilon_{9,j}, j \in \{1,\dots,4\}$
Dataset $\varepsilon_{10,j}, j \in \{1,\dots,4\}$
Probability density / A.U.
Energy / keV